\documentclass[prd,amsfonts,preprintnumbers,floats,tightenlines,floatfix,preprint,nofootinbib,showkeys,showpacs,superscriptaddress]{revtex4}
\usepackage{ifthen}                   % needed for Feyman-Slash macro
\usepackage{exscale}                  % correct scaling of math-symbols
\usepackage[intlimits]{amsmath}       % improved mathematical typesetting
\usepackage{amsfonts}
\usepackage{amssymb,amscd}
\usepackage{epsfig}
\newcommand{\pslash}{p\!\!\!/\,}
\newcommand{\qslash}{q\!\!\!/\,}

\newcommand{\cV}{\mathcal{V}}

\newcommand{\msing}{m_{\hbox{\scriptsize sing}}}
 
\begin{document}

\date{\today} \title{
  \hspace*{\fill}{\small\sf http://arXiv.org/abs/hep-ph/0309077}\\
  \hspace*{\fill}{\small\sf UNITU--THEP--12/03}\\
  \hspace*{\fill}{\small\sf NT@UW-03-22}\\
  \hspace*{\fill}{\small\sf HD-THEP-03-44}\\[4mm]

  Analytic properties of the Landau gauge\\ gluon and quark propagators}

\author{R.~Alkofer\footnote{E-Mail: reinhard.alkofer@uni-tuebingen.de}} 
\affiliation{Institute for Theoretical Physics, 
        University of T\"ubingen, D-72076 T\"ubingen, Germany}
  
\author{W.~Detmold\footnote{E-Mail: wdetmold@phys.washington.edu}}
\affiliation{Department of Physics, University of Washington, 
             %Box  351560, 
             Seattle WA 98195, U.S.A.}
\author{C.~S.~Fischer\footnote{Current address: Institute for Particle
    Physics Phenomenology, University of Durham, Durham DH1 3LE, U.K. 
    E-Mail: christian.fischer@durham.ac.uk}}
\affiliation{Institute for Theoretical Physics, 
        University of T\"ubingen, D-72076 T\"ubingen, Germany}
\affiliation{Institute for Theoretical Physics, 
        University of Heidelberg, D-69120 Heidelberg, Germany}
\author{P.~Maris\footnote{Current address: Department of Physics and
    Astronomy, University of Pittsburgh, Pittsburgh, PA 15260, U.S.A. 
    E-Mail: pim6@pitt.edu}}
\affiliation{
        Institute for Theoretical Physics, 
        University of T\"ubingen, D-72076 T\"ubingen, Germany}

\begin{abstract}
  We explore the analytic structure of the gluon and quark propagators
  of Landau gauge QCD from numerical solutions of the coupled system
  of renormalized Dyson--Schwinger equations and from fits to lattice
  data.  We find sizable negative norm contributions in the transverse
  gluon propagator indicating the absence of the transverse gluon from
  the physical spectrum.  A simple analytic structure for the gluon
  propagator is proposed.  For the quark propagator we find evidence
  for a mass-like singularity on the real timelike momentum axis, with
  a mass of 350 to 500 MeV.  Within the employed Green's functions
  approach we identify a crucial term in the quark-gluon vertex that
  leads to a positive definite Schwinger function for the quark
  propagator.
\end{abstract}

\maketitle

\section{Introduction}

Dynamical chiral symmetry breaking and confinement are fundamental
properties of QCD.  In high energy processes such as deep inelastic
scattering, quarks behave almost masselessly. However at low energies
the observed hadron spectrum suggests that light quarks acquire large,
dynamically generated masses through their interaction with the gauge
sector of QCD.  Quarks and gluons carry color charge and are not
observed as asymptotic states, only occurring inside colorless bound
states, the hadrons.  The mechanism for such confinement in QCD is
still not understood and it is not known whether a gauge invariant
formulation even exists.  However, in the framework of a quantum
theory, physical degrees of freedom are necessarily subject to a
probabilistic interpretation implying unitarity and positivity; the
physical part of the state space of QCD should be equipped with a
positive (semi-)definite metric.  Therefore one way to investigate
whether a certain degree of freedom is confined, is to search for
positivity violations in the spectral representation of the
corresponding propagator.  Negative norm contributions to the spectral
function signal the absence of asymptotic states from the physical
part of the state space of QCD and are thus a sufficient (though not
necessary) criterion for the confinement of the particle in question.

Neither confinement nor dynamical chiral symmetry breaking can be
accounted for at any finite order in perturbation theory.  These
phenomena can only be explored in genuinely non-perturbative
approaches such as those provided by lattice Monte--Carlo simulations
(see {\em e.g.} Ref.~\cite{Montvay:cy,Rothe:kp}) and the
Dyson--Schwinger, Green's functions approach (see {\em e.g.}
Refs.~\cite{Alkofer:2000wg,Roberts:2000aa,Maris:2003vk}).  Both
approaches have their own strengths and weaknesses.  Lattice
simulations are the only {\em ab initio} calculations available so
far.  They contain the full non-perturbative structure of QCD but are
limited by the enormous computational effort they require and by
uncertainties in the infinite volume and continuum extrapolations that
are needed to connect with the physical world.  Furthermore, the
implementation of small quark masses in most lattice simulations is
computationally very expensive and, as yet, state-of-the-art
calculations use light quark masses 6--10 times the physical values,
thus necessitating a further extrapolation.  On the other hand, the
Dyson--Schwinger equations for the propagators of QCD are
continuum-based and can be solved analytically in the infrared but
must be truncated to obtain a closed, solvable system of equations
\cite{vonSmekal:1997is,Atkinson:1998tu,Lerche:2002ep,Zwanziger:2001kw}.
Recently, a concerted effort has been made to combine the strengths of
these two approaches and quite definite statements on the infrared
behavior of QCD have
emerged~\cite{Fischer:2002hn,Fischer:2003rp,Tandy:2003hn,Bhagwat:2003vw}.
In this work we will apply a similar strategy to explore the analytic
structure of the propagators of QCD from solutions in the spacelike
Euclidean momentum region.

This paper is organized as follows: In Sec.~\ref{sec:pos} we briefly
review the connection between positivity and confinement and outline
the method we will use to investigate the analytic structure of the
propagator in the timelike momentum region.  In the third section we
investigate positivity violation in the gluon and quark propagators
which are obtained as solutions of Dyson--Schwinger equations in the
truncation scheme of Refs.~\cite{Fischer:2002hn,Fischer:2003rp}.  We
find clear evidence for positivity violations in the gluon propagator.
The origin of these positivity violations is a branch point at
$p^2=0$, followed by a cut along the real timelike axis.  For the
quark propagator we find no positivity violations as long as a certain
non-perturbative Dirac structure is included in the quark-gluon
vertex.  This Dirac structure is dictated by the Ward--Takahashi
identity in QED, and is also likely to exist in QCD because of the
similar nature of the corresponding Slavnov--Taylor identity.  In
Sec.~\ref{sec:quarkparam} we seek parameterizations of the quark
propagator.  We investigate the ability of a number of meromorphic
{\em ansaetze} to reproduce lattice data for the quark propagator.
All the fits share the property of either a dominant real pole or a
pair of complex conjugate poles very close to the real momentum axis.
We also show that one can reproduce both the Dyson--Schwinger
solutions and the lattice data by various parameterizations with
branch point singularities, rather than poles.  We give a summary of
our results in the last section.

\section{Positivity and Confinement \label{sec:pos} }

One of the most intricate problems in quantum field theories is the
separation of physical and unphysical degrees of freedom.  In QCD this
problem is directly connected with the issue of confinement, since we
are searching for the mechanism which eliminates the colored degrees
of freedom from the physical subspace, $\cV_{phys}$, of the state
space of QCD.  In order to ensure a probabilistic interpretation of
the quantum theory, $\cV_{phys}$ is required to be positive
semi-definite, whereas the total state space of QCD in covariant
gauges has an indefinite metric.

A possible definition of a positive definite subspace, $\cV_{phys}$,
is given in the framework of the Kugo--Ojima confinement scenario
\cite{Kugo:1979gm}.  Assuming the existence of a well-defined BRST
charge operator, $Q_B$, the space of physical states is defined by
\begin{equation} 
  {\cV_{phys}} = {\left\{|phys\rangle : Q_B|phys\rangle=0\right\}}.  
\end{equation}
Given the assumption of a well-defined, {\em i.e.\/} unbroken, global
color charge, $Q^a$, it has been shown that the physical state space
$\cV_{phys}$ only contains color singlets, {\em i.e.}  $\langle phys|
Q^a | phys \rangle=0$ \cite{Kugo:1979gm,Nakanishi:qm}.  In Landau
gauge this assumption, the Kugo--Ojima confinement criterion, can be
translated into the requirement that the ghost propagator should
diverge more strongly than a simple pole at zero momentum
\cite{Kugo:1995km}.

In this scenario, longitudinal gluons as well as ghosts are removed
from the physical spectrum of QCD by the BRST quartet mechanism (see
{\em e.g.} Ref.~\cite{Nakanishi:qm}).  The colored states are
BRST-quartet states, consisting of two parent and two daughter states
of respectively opposite ghost numbers.  The latter states are
BRST-exact and thus BRST-closed (due to the nilpotency of the BRST
transformation).  The BRST daughters are orthogonal to all other
states in the positive definite subspace and thus do not contribute to
physical $S$-matrix elements.  The parent states belong to the
indefinite metric part of the representation space and are thus
expected to violate positivity.  Members of the elementary quartet
related to gauge fixing are the ghosts, the antighosts and
longitudinal gluons.

As the two parent states of a quartet belong to the indefinite metric
part of the complete representation space, violation of positivity
would provide evidence for the correctness of the Kugo--Ojima picture.
{\em E.g.\/} positivity violation for transverse gluons indicates that
transverse gluons are BRST-parent states with gluon-ghost states as
daughters.  The corresponding parents of opposite ghost number are
gluon-antighost states with a mixture of gluon-ghost-antighost and
2-gluon states as daughters.  A similar construction for quarks would
consider quarks as BRST-parent states with quark-ghost states as
daughters, and correspondingly, quark-antighost states as second set of
parents and a mixture of quark-ghost-antighost and quark-gluon states
as second type of daughter states.  Thus an investigation of
(non-)positivity of transverse gluons and quarks allows us to
understand in more detail confinement via the BRST quartet mechanism

In order to complete the proof of confinement in this scenario one
must still demonstrate the appearance of a mass gap in $\cV_{phys}$
and the violation of cluster decomposition (see {\em e.g.}
Ref.~\cite{Strocchi:ci,Nakanishi:qm} and references therein) for
colored states.  Both requirements are related to the area law in the
Wilson loop and, correspondingly, to a non-vanishing string tension in
the quark-antiquark potential.
 
At this point we note that the basic assumption of the Kugo--Ojima
confinement scenario still seems far from being proved: BRST-symmetry
is a perturbative concept and it is not clear whether the symmetry
remains unbroken in non-perturbative QCD \cite{vanBaal:1997gu}.
Furthermore, although clear evidence for a linearly rising potential
between static quarks has been found in quenched lattice simulations
(see Ref.~\cite{Greensite:2003bk} and references therein), a
mathematical proof of a violation of cluster decomposition is not at
hand.  Nonetheless, the Kugo--Ojima confinement criterion in its
Landau gauge formulation has been tested in Dyson--Schwinger studies
and in lattice simulations.  Both methods agree very well even on a
quantitative level and find a strongly diverging ghost propagator at
small momenta
\cite{Suman:1996zg,Langfeld:2002dd,Furui:2003jr,Alkofer:2000wg,Fischer:2002hn,Fischer:2003rp}.

The Kugo--Ojima scenario is one particular mechanism that ensures the
probabilistic interpretation of the quantum theory.  However, even if
it were eventually shown not to be appropriate, it is apparent that
there is {\em some} mechanism which singles out a physical, positive
semi-definite subspace in QCD.  This suggests another criterion for
confinement, namely {\em violation of positivity}.  If a certain
degree of freedom has negative norm contributions in its propagator,
it cannot describe a physical asymptotic state, {\em i.e.\/} there is
no K\"all\'en--Lehmann spectral representation for its propagator.

Within the framework of a Euclidean quantum field theory (which is
used throughout this work) positivity is formulated in terms of the
Osterwalder--Schrader axiom of {\em reflection positivity}
\cite{Osterwalder:1973dx}.  (For a thorough mathematical formulation
of the axiom the reader is referred to
Refs.~\cite{Haag:1992hx,Glimm:ng}).  In the special case of a
two-point correlation function, $\Delta(x-y)$, the condition of
reflection positivity can be written as
\begin{equation} 
 \int d^4x \; d^4y \; 
 \bar{f}(\vec{x},-x_0) \; \Delta(x-y) \; {f}(\vec{y},y_0) \; \ge 0 \;,
\end{equation}
where $f(\vec{x},x_0)$ is a complex valued test function with support
in $\{(\vec{x},x_0) \: : \: x_0 > 0 \}$, {\em i.e.\/} for positive
times.  After a three-dimensional Fourier transformation, this
condition implies
\begin{equation} 
  \int_0^\infty dt \; dt^\prime \;
  \bar{f}(t^\prime,\vec{p}) \; \Delta(-(t+t^\prime),\vec{p}) \;
  f(t,\vec{p}) \; \ge 0\,. \label{pos-res}
\end{equation} 
Provided there is a region around $t_0=-(t+t^\prime)$ where
$\Delta(t_0,\vec{p}) <0$, one can easily find a real test function
$f(t)$ which peaks strongly at $t$ and $t^\prime$ and thereby
demonstrate positivity violation.  For the special case $\vec{p}=0$,
the Osterwalder--Schrader condition, Eq.~(\ref{pos-res}), can be given
in terms of the Schwinger function, $\Delta(t)$, defined by
\begin{equation} 
  \Delta(t) := \int d^3x \int \frac{d^4p}{(2\pi)^4}
  e^{i(t p_4+\vec{x}\cdot\vec{p})} \sigma(p^2) 
  = \frac{1}{\pi}\int_0^\infty
  dp_4 \cos(t \; p_4) \sigma(p^2_4) \;\ge 0 \,,
\label{schwinger} 
\end{equation}
where $\sigma(p^2)$ is a scalar function extracted from the
corresponding propagator.  For the propagator of transverse gluons,
$\sigma(p^2)$ is simply given by the renormalization function times
the tree-level expression $1/p^2$ (see Eq.~(\ref{gluon_prop}) below)
and we denote the corresponding Schwinger function by $\Delta_g(t)$.
The quark propagator can be decomposed into a scalar and a vector part
\begin{eqnarray}
  S(p) & =: &  i \pslash \sigma_v(p^2) + \sigma_s(p^2) \;,
\end{eqnarray}
leaving us with two scalar functions, $\sigma_{v}(p^2)$ and
$\sigma_{s}(p^2)$, to form two Schwinger functions, $\Delta_v(t)$ and
$\Delta_s(t)$.

Two simple examples for the analytic structure of a propagator in a
quantum field theory are a real pole and a pair of complex conjugate
poles.  These highlight the paradigmatic behaviors of the Schwinger
function, Eq.~(\ref{schwinger}).  In the following, we always discuss
the propagators and the functions $\sigma_{s,v}(p^2)$ in terms of the
Lorentz invariant complex momentum, $p^2$.  Our notation is such that
positive real values, $p^2>0$, correspond to spacelike momenta.

(I) {\em Real pole}.  
%%%
The propagator of a real, massive, scalar particle has a single pole
on the real timelike ($p^2<0$) momentum axis.  In this case the
propagator function is given by $\sigma(p^2)=1/(p^2+m^2)$ and it is
easy to see from Eq.~(\ref{schwinger}) that the Schwinger function
decays exponentially,
\begin{equation}
  \Delta(t) \sim e^{-m\;t}\,, 
\label{decay} 
\end{equation} 
and is positive definite.  For a bare propagator, the pole mass, $m$,
is the same as the bare mass occurring in the Lagrangian.  However,
for an interacting particle, the pole mass can have both tree level
and dynamically generated contributions.  The real pole corresponds to
the presence of a stable asymptotic state associated with this
propagator.  This does not imply that this state corresponds to an
observable physical particle: provided the Kugo--Ojima scenario holds,
all states belonging to a quartet representation of the BRST-algebra
are excluded from the physical subspace, $\cV_{phys}$, which contains
only colorless singlets.  Thus two-point correlations of colored
fields may develop real poles in momentum space without contradicting
confinement~\cite{Oehme:1994pv}.  In lattice calculations
\cite{Montvay:cy} and other non-perturbative approaches
\cite{HollenbergBurkardt}, the exponential decay in Eq.~(\ref{decay})
is used to extract hadron masses and other observables from the large
time behavior of appropriate correlators.

(II) {\em Complex conjugate poles.}  
%%%
Another possible analytic structure for a propagator is a pair of
complex conjugate poles with ``masses'' $m=a\pm i b$.  As has been
discussed in detail in Refs.~\cite{Stingl:1996nk}, such a propagator
could describe a short lived excitation which decays exponentially at large
timelike distances.  Furthermore, it has been argued
\cite{Stingl:1996nk} that although causality is violated at the level
of the propagators, the corresponding S-matrix remains both causal and
unitary.  Such complex conjugate poles lead to oscillatory behavior
in the Schwinger function, $\Delta(t)$.  Specifically,
\begin{equation} 
  \Delta(t) \sim e^{-a\;t} \: \cos(bt + \delta) \,.
  \label{oszillation} 
\end{equation} 
In this case one has negative norm contributions to the Schwinger
function and the effective mass, 
\begin{equation} 
 m_{\rm eff}(t) = -\frac{d\ln\Delta(t)}{d t}
\end{equation} 
(defined in analogy to the real pole case, Eq.~(\ref{decay})) exhibits
periodic singularities.  Therefore the associated state (if there is
any) must be an element of the unphysical subspace.  Under the
assumption of an unbroken BRST symmetry, this state must be a member
of a BRST quartet, and the corresponding excitation is confined.

Complex conjugate poles have been found for the fermion propagators of
QED$_3$ \cite{Maris:1995ns}, QED$_4$ (see {\em e.g.}
\cite{Atkinson:1978tk}), and QCD
\cite{Bhagwat:2003vw,Stainsby:1990fh,Maris:1991cb,Bender:1996bm,Bender:1997jf,Burden:1997ja}
in a variety of truncation schemes.  In a number of these studies, the
authors have discussed whether the observed positivity violations are
genuine properties of the theory related to confinement or artifacts
of the truncation schemes
\cite{Maris:1995ns,Stainsby:1990fh,Krein:1990sf,Burden:1991gd}.  As
examined in the following section, it is our contention that dominant
complex conjugate poles are indeed an artifact of the rainbow (bare
vertex) truncation of the quark Dyson--Schwinger equation and that, at
least in Landau gauge, confinement through positivity violation in the
quark propagator is not manifest.  Complex conjugate propagators are
also known to be practicable in light-cone dominated processes
\cite{Tiburzi:2003ja} and have recently been investigated in terms of
the solution of the Bethe--Salpeter equation \cite{Bhagwat:2003wu}.
It has also been suggested that the gluon propagator may have such an
analytic structure
\cite{Stingl:1996nk,Gribov:1978wm,Zwanziger:1991ac,Driesen:1997wz}.
This possibility has been investigated in
Refs.~\cite{Hawes:ef,Bender:1994bv}.

Here, a note on positivity for the propagator of a Dirac field is in
order.  A dispersion relation representation of a fermion propagator
in Minkowski space reads
\begin{equation}
 S(p) = \int_0^\infty ds \frac{\pslash \rho_v(s) + \rho_s(s)}
 {p^2-s +i \epsilon}   \;,
\end{equation}
and positivity amounts to the requirements that for $s>0$
\begin{equation}
\rho_v(s) \ge 0 \;, \qquad 
{\rm and} \qquad \sqrt{s} \; \rho_v(s) - \rho_s(s) \ge 0 \;.
\label{eq:dispersionconstraint}
\end{equation}
It is obvious that for a free Dirac field of mass $m$ one has
\begin{equation}
\rho_v(s) = \delta (s-m^2) \;,\qquad 
{\rm and} \qquad \rho_s(s) = m \; \delta (s-m^2) \;,
\end{equation}
and thus $\sqrt{s} \rho_v(s) - \rho_s(s) =0$.  For an interacting
Dirac field with physical asymptotic states and mass $m$ one expects
$\rho_{s,v}(s) = 0$ for $s < m^2$.  For $s>m^2$,
Eq.~(\ref{eq:dispersionconstraint}) has to be satisfied.  This
requirement is automatically fulfilled if the stronger constraint 
\begin{equation}
  m\rho_v(s) \ge \rho_s(s)  \,,
\label{eq:spectralvsconstraint}
\end{equation}
holds.

Given the linearity of the different types of integral transforms
relating $\sigma_{s,v}(p^2)$, $\rho_{s,v}(s)$, and $\Delta_{s,v}(t)$
to each other, one can conclude that $\sigma_v(p^2)$ must be
multiplied by a typical mass scale before being compared to
$\sigma_s(p^2)$.  Thus, positivity violations can be signaled either
in $\sigma_v(p^2)$ alone, or in appropriate linear combinations of
$\sigma_v(p^2)$ and $\sigma_{s}(p^2)$.  We also consider the Schwinger
function associated solely with $\sigma_s(p^2)$, since it can be
calculated with greater numerical accuracy.  In general, oscillatory
behavior in $\Delta_s(t)$ signals oscillatory behavior in
$\Delta_v(t)$ as well.

Using the corresponding Schwinger functions, we can search for
possible positivity violations and investigate the analytic structure
of the gluon and quark propagators of QCD.  The $t$-dependencies of
these Schwinger functions are determined by the analytic properties of
the propagator, and, for large $t$, are dominated by the singularity
closest to $p^2 = 0$.  A complementary, direct method of determining
the analytic structure is to solve the corresponding Dyson--Schwinger
equation over a large region of the complex momentum plane.  However,
from a numerical point of view, such a procedure is very expensive and
is not feasible with the resources currently available to us.
Furthermore, there is good evidence from an investigation of QED$_3$
that both methods agree very well \cite{Maris:1995ns}.  We are thus
confident that the Fourier transformation method is able to determine
the qualitative behavior of the propagators.

To complete this discussion we note that the conversion of a
tree-level pole into an algebraic branch point with exponent larger
than one is also known for certain approximations to the fermion
propagator of QED$_4$ (see, {\em e.g.}, supplement S4 in
Ref.~\cite{Jau76} and references therein). This type of singularity,
$(p^2+m^2)^{-1-\alpha /\pi}$, is related to the soft photon cloud.
The examples discussed in this section (real poles, complex conjugate
poles, or branch cuts) will form the basis of our investigation of the
analytic structure for the quark and gluon propagators.

\section{Solutions of the propagator Dyson--Schwinger equations
of Landau gauge QCD \label{sec:DSEsolns} } 

In this section we present solutions of the coupled set of
Dyson--Schwinger equations (DSEs) for the ghost, gluon, and quark
propagators in Landau gauge and investigate some of their analytic
properties.  In order to keep this paper self-contained, we first
briefly review the DSE truncation scheme developed in
Refs.~\cite{Fischer:2002hn,Fischer:2003rp} which is used to determine
the propagators for Euclidean spacelike momenta, i.e. for real $p^2\ge
0$.  It is important to note that the behavior of the propagators for
$p^2\to 0^+$ is extracted analytically.

The DSEs for the quark, gluon and ghost propagators are derived from
the QCD generating functional with gluon field configurations
restricted to the first Gribov region \cite{Gribov:1978wm}.  In a
recent work it has been argued that such a prescription is sufficient
to eliminate the effects of Gribov copies from correlation functions
\cite{Zwanziger:2003cf}.  Furthermore, the DSEs are not affected by
imposing such a boundary condition on the generating functional of the
gauge fixed theory because the Gribov horizon is a nodal surface for
the integrand of this functional integral.  Instead, the ghost
two-point function has to satisfy the so-called horizon condition
\cite{Zwanziger:2001kw}, {\em i.e.} the ghost propagator has to
diverge more strongly than a simple pole for $p^2\to 0^+$.  This
condition (which in Landau gauge is formally equivalent to the
Kugo--Ojima confinement criterion discussed in the preceding section)
turns out to be enforced by the ghost DSE
\cite{Watson:2001yv,Alkofer:2001iw,Lerche:2002ep} and is thus
fulfilled by the DSE solutions in the truncation scheme that we
employ.

%%%%%%%%%%%%%%%%%%%%%%%%%%%%%%%%%%%%%%%%%%%%%%%%%%%%%%%%%%%%%%%%%%%%%%%%%%%%
\begin{figure}[t!]
  \centerline{
    \epsfig{file=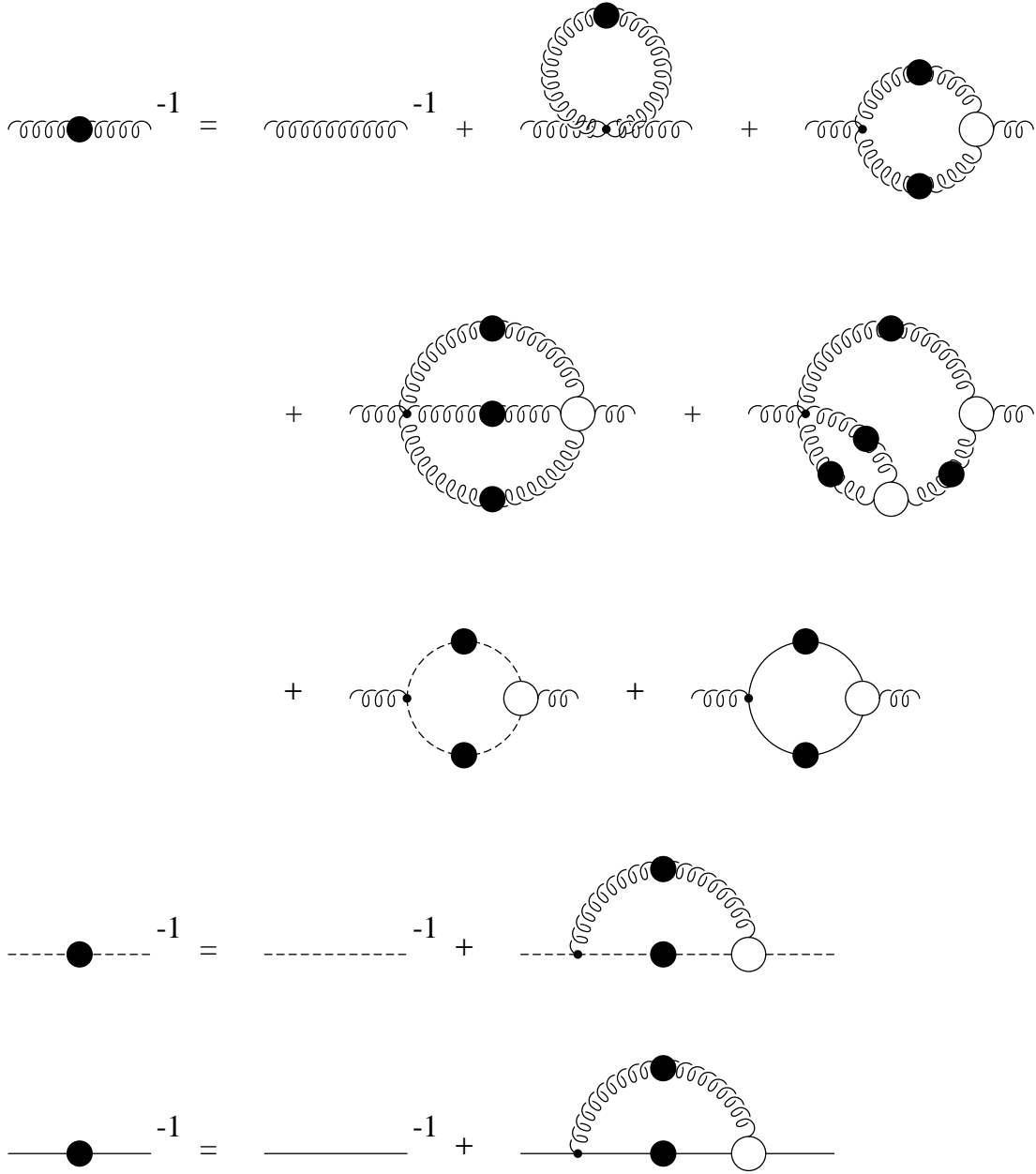,width=0.9\linewidth}%,height=15cm}
  } \vskip 5mm
  \caption{\label{GluonGhostQuark}
    Diagrammatic representation of the Dyson--Schwinger equations for 
    the gluon, ghost, and quark propagators.  The wiggly, dashed, and 
    solid lines represent the propagation of gluons, ghosts, and quarks, 
    respectively.  A filled blob represents a full propagator and 
    a circle indicates a one-particle irreducible vertex.} 
\end{figure}
%%%%%%%%%%%%%%%%%%%%%%%%%%%%%%%%%%%%%%%%%%%%%%%%%%%%%%%%%%%%%%%%%%%%%%%%%%%
A graphical representation of the DSEs for the ghost, gluon, and quark
propagators is given in Fig.~\ref{GluonGhostQuark} and their full form
can be found in Ref.~\cite{Alkofer:2000wg}.  In Landau gauge (which is
used throughout this work), the renormalized ghost, gluon and quark
propagators, $D_G(p,\mu)$, $D_{\mu\nu}(p,\mu)$, and $S(p,\mu)$,
respectively, are given in terms of scalar functions by
\begin{eqnarray}
  D_G(p,\mu) &=& - \frac{G(p^2,\mu^2)}{p^2} \,, 
  \label{ghost_prop}
\\
  D_{\mu \nu}(p,\mu) &=& \left(\delta_{\mu \nu} - \frac{p_\mu
      p_\nu}{p^2} \right) \frac{Z(p^2,\mu^2)}{p^2} \,,
  \label{gluon_prop} 
\\
  S(p,\mu) &=& \frac{1}{-i \pslash A(p^2,\mu^2) + B(p^2,\mu^2)} \:\:\:
  =: \:\:\: i \pslash \sigma_v(p^2,\mu^2) + \sigma_s(p^2,\mu^2)
  \, . 
  \label{quark_prop} 
\end{eqnarray}
All these propagators are diagonal in their respective representations
of SU($N_c$), so their color structure has been suppressed
for simplicity.  The dependence on the renormalization scale, $\mu$,
is given explicitly for later use.  Here, $G(p^2,\mu^2)$ and
$Z(p^2,\mu^2)$ are the ghost and gluon dressing functions,
respectively, and $A(p^2,\mu^2)$ and $B(p^2,\mu^2)$ are the vector and
the scalar parts of the inverse of the quark propagator.  The functions
most relevant for our study of positivity are $Z(p^2,\mu^2)/p^2$,
$\sigma_{s}(p^2,\mu^2)$ and $\sigma_{v}(p^2,\mu^2)$.  Note that the
ghost propagator trivially violates reflection positivity because of
the way ghosts are introduced in Faddeev--Popov quantization
\cite{Faddeev:fc}.

Two renormalization-scale-independent combinations built from the
scalar functions representing the different propagators are important
for further discussion.  First, $M(p^2)=B(p^2,\mu^2)/A(p^2,\mu^2)$
denotes the renormalization-point-independent quark mass function.
Second, as has been demonstrated in Ref.~\cite{vonSmekal:1997is}, a
non-perturbative definition of the running coupling, is possible due
to the non-renormalization of the ghost-gluon vertex in Landau gauge \cite{Taylor:ff}.
This results in the relation
\begin{equation} 
  \alpha(p^2) =
  \alpha(\mu^2) \: G^2(p^2,\mu^2) \: Z(p^2,\mu^2) \;. 
  \label{alpha_def}
\end{equation} 
In the following we investigate the full ({\em unquenched}) system of
DSEs and also the {\em quenched} approximation to them in which quark
loops are neglected, removing the back-reaction of the quarks on the
ghost and gluon system.

\subsection{Truncation scheme}

Both the quenched and the unquenched system of ghost, gluon, and quark
DSEs have been solved numerically in
Refs.~\cite{Fischer:2002hn,Fischer:2003rp} in a truncation scheme
which neglects the effects of the four-gluon interaction and employs
{\em ansaetze} for the ghost-gluon and the three-gluon vertices such
that two important constraints are fulfilled: the running coupling,
$\alpha(p^2)$, is independent of the renormalization point and the
anomalous dimensions of the ghost and gluon propagators are reproduced
at the one-loop level for large momenta.  In order to study the
effects of violating gauge invariance by these truncation assumptions,
the gluon DSE has been contracted with the one-parameter family of
tensors
\begin{equation}
  {\mathcal P}^{(\zeta )}_{\mu\nu} (p) = \delta_{\mu\nu} - 
  \zeta \,\frac{p_\mu p_\nu}{p^2} \, . 
  \label{Paproj}
\end{equation} 
In Landau gauge, a violation of gauge invariance manifests itself in
the appearance of spurious longitudinal terms in the gluon equation,
which in turn introduces dependence of the ghost and gluon dressing
functions on the parameter $\zeta$.  The influence of these
longitudinal terms has been examined in Ref.~\cite{Fischer:2002hn} by
varying $\zeta$ and found to be surprisingly small.  Further technical
details of the truncation scheme in the Yang-Mills sector are
relegated to Appendix A where we also discuss the dependence of our
analysis on these details (see also
Refs.~\cite{Fischer:2002hn,Fischer:2003rp}).

Employing asymptotic expansions for the propagators at small momenta,
the untruncated ghost and gluon DSEs can be solved analytically for
$p^2\to0^+$ \cite{Watson:2001yv}.  One finds simple power laws, with
exponents related as
\begin{eqnarray}
  Z(p^2,\mu^2) &\sim& (p^2/\mu^2)^{2\kappa} \; , 
  \label{z-power}
\\
  G(p^2,\mu^2) &\sim& (p^2/\mu^2)^{-\kappa} \; , 
  \label{g-power} 
\end{eqnarray}
for the gluon and ghost dressing functions.  The value of the exponent
$\kappa$ depends somewhat on the details of the employed truncation
scheme. In certain truncations it can be calculated analytically and
it will depend on the parameter $\zeta$ \cite{Fischer:2002hn}.  The
tensor ${\mathcal P}^{(\zeta=1)}_{\mu\nu}$ projects onto the purely
transverse part of the gluon equation, and in this case the solution
$\kappa=(93-\sqrt{1201})/98 \approx 0.595$ has been found in
Refs.~\cite{Lerche:2002ep,Zwanziger:2001kw}. By varying $1 \le \zeta
<4$, infrared solutions with exponents in the range $0.5 < \kappa \leq
(93-\sqrt{1201})/98 $ have been shown to connect to numerical
solutions for all momenta \cite{Fischer:2002hn}.  A recent infrared
analysis of the ghost and gluon DSEs employing the most general {\em
  ansatz} for the ghost-gluon vertex suggests the exponent $\kappa$ is
in the range $0.5 < \kappa < 1$ \cite{Lerche:2002ep} (which is further
restricted to $0.5 < \kappa<0.7$ after constraints on the value of the
running coupling are taken into account). A first attempt to include
the two-loop diagrams in the gluon DSE also results in very similar
values for the infrared exponent \cite{Bloch:2003yu} and in
Ref.~\cite{Zwanziger:2003cf} it has been shown that the two-loop
diagrams have no effect on $\kappa$.  Finally, exact renormalization
group equations have recently been employed in a complementary
investigation \cite{Pawlowski:2003hq} of the infrared behavior of the
gluon and ghost propagators with a resulting value for $\kappa$ in
agreement with those above.  These varied investigations all indicate
that the Landau gauge gluon propagator vanishes as $p^2\to0^+$ and
predict an exponent $0.5 < \kappa < 0.7$.

For the subsequent discussion, it is important to note that the
exponent $\kappa$ is very likely an irrational number.  The relation
of the exponents in Eqs.~(\ref{z-power}) and (\ref{g-power}) results
in an infrared finite strong coupling independent of the value of
$\kappa$, {\em c.f.\/} Eq.~(\ref{alpha_def}).  For transverse
projection, the value is given by $\alpha(0) = 8.915/N_c$.

The DSE for the quark propagator $S(p,\mu)$ is given by 
\begin{equation} 
  S^{-1}(p,\mu) = Z_2(\mu^2,\Lambda^2) \, S^{-1}_0(p) + \frac{g^2}{16\pi^4}\, 
  Z_{1F}(\mu^2,\Lambda^2)\, C_F\, \int^\Lambda d^4q\,
  \gamma_{\mu}\, S(q,\mu) \,\Gamma_\nu(q,p;\mu) \,D_{\mu \nu}(k,\mu) \,,
  \label{quark1}
\end{equation} 
where $Z_2$ and $Z_{1F}$ are the quark wave function- and quark-gluon
vertex-renormalization constants, respectively, and $\int^\Lambda$
represents a translationally-invariant regularization characterized by
a scale, $\Lambda$.  The momentum routing is $k=q-p$, and the factor
$C_F=(N_c^2-1)/2N_c$ stems from the color trace of the loop.

In addition to the quark and gluon propagators, Eq.~(\ref{quark1})
involves the quark-gluon vertex, $\Gamma_\nu(q,p;\mu)$.  This vertex
is, in principle, determined by its own DSE \cite{Roberts:dr}
involving various ($n\le5$)-point correlators.  However, the solution
of such higher-order DSEs is difficult even in the simplest situations
\cite{Detmold:2003au} and we avoid the problem by making an {\em
ansatz} for $\Gamma_\nu(q,p;\mu)$.  As the structure of this vertex
turns out to be crucial in our analysis of positivity violations in
the quark propagator, we explore its construction in some detail.

A reasonable {\em ansatz} for the quark-gluon vertex has to satisfy at
least two constraints: it should guarantee the multiplicative
renormalizability of the quark propagator in the quark DSE, and it
should at least approximately satisfy its non-Abelian Slavnov--Taylor
identity.  It has been shown in Ref.~\cite{Fischer:2003rp} that the
construction
\begin{equation}
  \Gamma_\nu(q,p;\mu) = V_\nu^{abel}(q,p;\mu) \, 
                        W^{\neg abel}(p^2,q^2,k^2;\mu) \;,
  \label{vertex-ansatz}
\end{equation}
with
\begin{eqnarray} 
  W^{\neg abel}(p^2,q^2,k^2;\mu) &=& G^2(k^2,\mu^2)
  \:\tilde{Z}_3(\mu^2,\Lambda^2) \; ,
  \label{vertex_nonabel}
\\
  V^{abel}_\nu(q,p;\mu) &=& \Gamma_\nu^{CP}(q,p;\mu) \nonumber \\ &=&
  \frac{A(p^2,\mu^2)+A(q^2,\mu^2)}{2} \gamma_\nu + i
  \frac{B(p^2,\mu^2)-B(q^2,\mu^2)}{p^2-q^2} (p+q)_\nu
  \nonumber\\
  && {} + \frac{A(p^2,\mu^2)-A(q^2,\mu^2)}{2(p^2-q^2)}
  (\pslash+\qslash)(p+q)_\nu
  \nonumber\\
  && {} + \frac{A(p^2,\mu^2)-A(q^2,\mu^2)}{2} \left[(p^2-q^2)\gamma_\nu -
    (\pslash-\qslash)(p+q)_\nu \right] \nonumber\\
  && \hspace*{3.5cm} \times
  \frac{p^2+q^2}{(p^2-q^2)^2+(M^2(p^2)+M^2(q^2))^2} \;,
  \label{vertex_CP}
\end{eqnarray} 
and $\tilde{Z}_3$ being the ghost wave function renormalization
constant, satisfies these requirements.  Here it is assumed that the
non-Abelian part of the vertex, $W^{\neg abel}(p^2,q^2,k^2;\mu)$, can
be factored out from the Dirac structure, and that the Dirac structure
is given by $\Gamma_\nu^{CP}(q,p;\mu)$, the Curtis--Pennington (CP)
construction of the fermion-photon vertex in QED$_4$
\cite{Curtis:1990zs,Ball:ay}.  Note that the dressing of the
longitudinal part of the CP vertex is dictated by the Abelian Ward
identity
\begin{equation}
  -i k_\mu \: \Gamma_\mu^{QED}(q,p;\mu) = 
                      S^{-1}(p,\mu) -  S^{-1}(q,\mu) \;,
  \label{QED-quark-gluon-STI}
\end{equation}
which results, among other things, in the appearance of a quark-gluon
coupling term proportional to the sum of the incoming and outgoing
quark momenta,
\begin{equation} 
  \Delta B_\nu := i\frac{B(p^2,\mu^2)-B(q^2,\mu^2)}{p^2-q^2} (p+q)_\nu \, . 
  \label{DeltaB} 
\end{equation} 
Such a coupling, being effectively scalar, may at first sight appear
to violate chiral symmetry, as, in contrast to the perturbatively
dominant vector coupling proportional to $\gamma_\nu$, the expression
(\ref{DeltaB}) commutes with $\gamma_5$.  However, this scalar term
only appears if chiral symmetry is already dynamically broken and is
thus consistent with the chiral Ward identities.  Its existence
provides significant additional (self-consistent) enhancement of
dynamical chiral symmetry breaking.  Such a scalar coupling also
appears in vertices that occur in systematic improvements on the
rainbow (bare vertex) truncation
\cite{Bender:2002as,Bender:1996bb,Hellstern:1997nv}.  This term will
be important in our investigations of positivity below.

For comparison, we also employ a construction with a bare Abelian part
of the vertex given by
\begin{eqnarray}
  V^{abel}_\nu(p,q;\mu) &=& Z_2(\mu,\Lambda) \: \gamma_\nu\,.
  \label{vertex_bare}
\end{eqnarray}
In both cases the input from the Yang-Mills sector of the theory, {\em
i.e.\/} the factors from the dressed gluon propagator and the
non-Abelian vertex dressing can be combined to give the running
coupling $\alpha (k^2) = g^2(\mu)\:G^2(k^2,\mu^2)\:Z(k^2,\mu^2)/4\pi$
according to Eq.~(\ref{alpha_def}).  Thus we arrive at the truncated
quark DSE
\begin{equation} 
  S^{-1}(p,\mu) = Z_2(\mu^2) \, S^{-1}_0(p) + 
  \frac{Z_2(\mu^2)}{3\pi^3}\, \int d^4q\, \frac{\alpha(k^2)}{k^2} 
  \left(\delta_{\mu\nu}-\frac{k_\mu k_\nu}{k^2}\right)
  \gamma_{\mu}\, S(q,\mu) \,V_\nu^{abel}(q,p;\mu) \,.
\label{eq:truncDSEquark}
\end{equation} 
In the quenched and unquenched calculations of the quark propagator we
take $\alpha(k^2)$ directly from the ghost and gluon equations.

We also consider the solutions of the quark DSE in the model
calculations of Refs.~\cite{Maris:1997tm,Maris:1999nt,Alkofer:2002bp}.
There, only the leading $\gamma_\mu$-part of the quark gluon vertex
has been employed and the combination of the gluon and vertex dressing
needed in the quark DSE has been modeled phenomenologically.  With
$\gamma_m = 12/(11 N_c - 2 N_f)$ being the anomalous dimension of the
quark propagator, we follow the authors of Ref.~\cite{Maris:1999nt}
and use the model
\begin{equation}
  \frac{\alpha(q^2)}{q^2} = \frac{\pi}{\omega^6}D\,q^2 \, e^{-q^2/\omega^2} 
   + \frac{\pi\,\gamma_m\; [1-\exp(-q^2/m_t^2)]}{q^2\;\frac{1}{2}
  \ln\left[e^2-1+(1+q^2/\Lambda^2_{QCD})^2\right]}\,,
  \label{model1} 
\end{equation} 
with $\Lambda_{QCD}=0.234\;{\rm GeV}$ in the $\overline{\mbox{MS}}$-scheme,
$N_f=4$ and the parameters $m_t=1.0\;\rm{GeV}$, $\omega=0.3\;\rm{GeV}$, and
$D=0.781\;\rm{GeV}^{2}$ fixed by fitting the chiral condensate and
pion decay constant.  Omitting the perturbative logarithmic tail, we
also compare with the model of Ref.~\cite{Alkofer:2002bp}, using a
purely Gaussian interaction
\begin{equation} 
  \frac{\alpha(q^2)}{q^2} = \frac{\pi}{\omega^6}D\,q^2 \,e^{-q^2/\omega^2} \,,
  \label{model2}
\end{equation} 
with $\omega=0.5\;\rm{GeV}$ and $D=1\;\rm{GeV}^{2}$.  

Despite the fact that these models for the effective interaction were
designed to be used in combination with a bare vertex, we also use
them in conjunction with the CP vertex, $\Gamma_\nu^{CP}$.  By
comparing quark propagators that result from employing either direct
input from the ghost and gluon sector or the model forms,
Eqs.~(\ref{model1}) and (\ref{model2}), we are in a position to test
whether the analytic properties of the quark propagator are more
sensitive to the global strength of the quark-gluon interaction, to
the overall shape of the (effective) running coupling, or to the
details of the tensor structure of the quark-gluon vertex.  First
however, we will discuss the results of the numerical calculations for
the gluon propagator.

\subsection{Results for the gluon propagator for Euclidean momenta}

%%%%%%%%%%%%%%%%%%%%%%%%%%%%%%%%%%%%%%%%%%%%%%%%%%%%%%%%%%%%%%%%%%%%%%%%%%
\begin{figure}[th!]
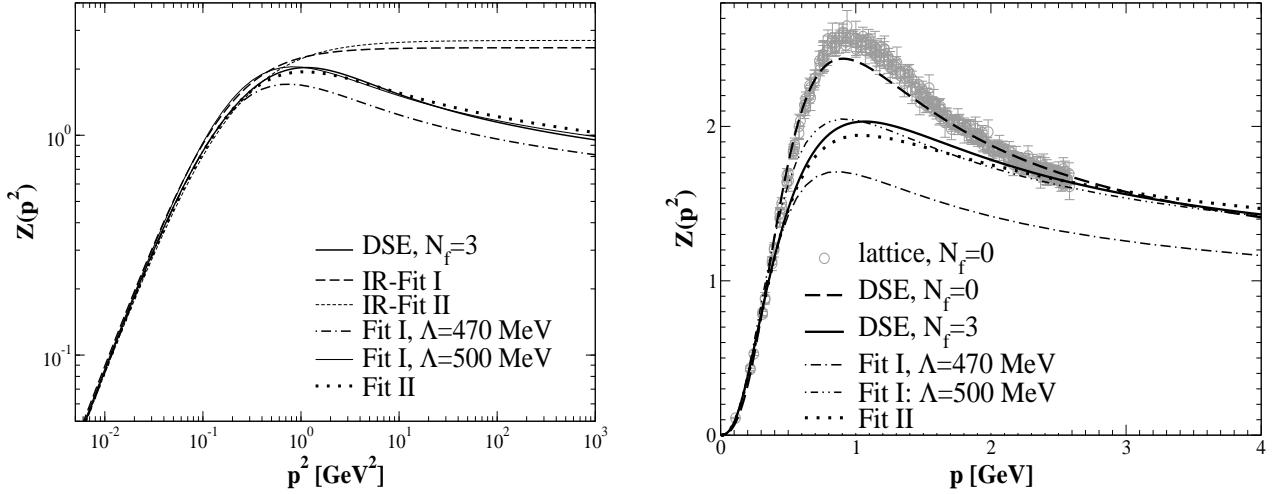

  \vspace{5mm} 
   \centerline{
    \epsfig{file=p.gluon2.eps,width=80mm,height=65mm}
    \hspace{5mm}
    \epsfig{file=p.gluon_latt2.eps,width=80mm,height=65mm} }
\caption{\label{gluon.dat}
  The solutionis of the quenched ($N_f=0$) and unquenched ($N_f=3$) 
  coupled DSEs for the gluon dressing function, $Z(p^2)$, are shown.
  The unquenched case with three massless flavors is compared to
  different fits (see text for details of the fits).  In the left
  panel these are displayed on logarithmic scales, in the right panel, 
  on linear scales.  Results from quenched lattice 
  calculations \cite{Bonnet:2001uh} are given in the right panel.}
\end{figure}
%%%%%%%%%%%%%%%%%%%%%%%%%%%%%%%%%%%%%%%%%%%%%%%%%%%%%%%%%%%%%%%%%%%%%%%%%%%%
 
In Fig.~\ref{gluon.dat} we display the numerical results for the gluon
dressing function calculated with zero (quenched) or three
(unquenched) flavors of massless quarks and transverse projection,
$\zeta=1$ ({\em c.f.\/} Eq.~(\ref{Paproj})), taken from
Ref.~\cite{Fischer:2003rp}.\footnote{As can be inferred from Refs.\ 
  \cite{Fischer:2002hn,Fischer:2003rp}, changing the projection of the
  gluon equation in the range $1\le \zeta < 4$ leads only to
  quantitative changes in the gluon and ghost renormalization
  functions.}  In the diagram on the right of Fig.~\ref{gluon.dat},
the DSE results are compared to results from quenched lattice
Monte--Carlo simulations \cite{Bonnet:2001uh}. The quenched DSE
results are seen to agree well with the lattice data. In contrast, the
unquenched DSE gluon propagator is significantly suppressed in the
intermediate momentum region where the screening effects of
quark-antiquark pairs become important.  For both $N_f=0$ and $3$,
there are two qualitative properties that we can extract from these
results: the analytically calculated infrared behavior given by
Eq.~(\ref{z-power}), and a maximum around $\sim 1\;\rm{GeV}$, followed
by relatively flat momentum dependence above this scale.

The behavior of the gluon dressing function in the infrared is
captured by either of the irrational functions\footnote{From here 
on we shall suppress the renormalization scale dependence (whenever
possible) for concision.}
\begin{eqnarray}
  Z^{ir}_I(p^2) &=& w_I \: \frac{(p^2)^{2 \kappa}}{(\Lambda^2_I)^{2
      \kappa} + (p^2)^{2 \kappa}} \,, 
  \label{IRfitI}\\
  Z^{ir}_{II}(p^2) &=& w_{II} \: \left(\frac{p^2}{\Lambda^2_{II} +
      p^2}\right)^{2 \kappa} \,, 
  \label{IRfitII} 
\end{eqnarray} 
which are exact in the infrared limit ({\em c.f.\/}
Eq.~(\ref{z-power})) and which play a role when it comes to the
interpretation of our results for the gluon Schwinger function,
$\Delta_g(t)$.  The value for the exponent $\kappa \simeq 0.595$ in
these fits is taken from the infrared analysis of the DSEs.  Note that
for $\kappa \to 1$, the form of Eq.~(\ref{IRfitI}) becomes identical
to the Gribov form proposed in
Refs.~\cite{Gribov:1978wm,Zwanziger:mf}.  The normalization parameters
$w_I$, $w_{II}$ and scales $\Lambda_I$, $\Lambda_{II}$ are chosen such
that the Schwinger function of the ($\zeta=1,\,N_f=3$) numerical gluon
propagator is reproduced by the Fourier transforms of the fits (the
value of these parameters are given below).  Our fits with these
irrational functions $Z^{ir}_{I,II}(p^2)$ are shown in
Fig.~\ref{gluon.dat} and clearly reproduce the behavior of the DSE
gluon propagator for very small momenta but deviate significantly from
the dressing functions at momenta above $\sim400$ MeV.

To describe the behavior for larger momenta, we multiply the functions
$Z^{ir}_{I,II}(p^2)$ by a function incorporating the known ultraviolet
behavior.  To this end we note that in Ref.~\cite{Fischer:2003rp} the
numerical running coupling has been fitted by\footnote{In
Ref.~\cite{Fischer:2003rp} two additional parameters $a$ and $b$ were
used with $a=1.020$ and $b=1.052$.  As the deviations from unity are
completely insignificant we have fixed $a=b=1$ here.}
\begin{equation}
\alpha_{\rm fit}(p^2) =  
\frac{\alpha(0)}{1+ p^2/\Lambda^2_{QCD}}
 + \frac{4 \pi}{\beta_0} \; \frac{p^2}{p^2 + \Lambda^2_{QCD}} \;
      \left(\frac{1}{\ln(p^2/\Lambda^2_{QCD})}- 
            \frac{\Lambda_{QCD}^2}{p^2-\Lambda_{QCD}^2}\right) \; .
\label{fitB}
\end{equation}
In this expression the Landau pole has been subtracted as has been
suggested in the framework of analytic perturbation theory
\cite{Shirkov:1997wi}.  The value $\alpha(0)=8.915/N_c$ is known from
the infrared analysis and $\beta_0 = (11 N_c - 2 N_f)/3$.  Using a MOM
scheme and fitting only the ultraviolet behavior, a value
$\Lambda_{QCD}=0.71\,\mbox{GeV}$ has been given in
Ref.~\cite{Fischer:2003rp}.

Identifying $\Lambda_{I,II}=\Lambda_{QCD}$ for simplicity, we utilize
the fits
\begin{eqnarray}
  Z_{I,\,II}(p^2) &=& Z^{ir}_{I,\,II}(p^2) \; 
           \alpha^{-\gamma}_{\rm fit}(p^2)\;,
  \label{fullfit}  
\end{eqnarray} 
for further investigations, using the one-loop value of the gluon
anomalous dimension, $\gamma = (-13 N_c + 4 N_f)/(22 N_c - 4 N_f)$.
The quality of these fits can be seen in Fig.~\ref{gluon.dat}.  For a
discussion of the parameters used, see below.

%%%%%%%%%%%%%%%%%%%%%%%%%%%%%%%%%%%%%%%%%%%%%%%%%%%%%%%%%%%%%%%%%%%%%%%%%%
\begin{figure}[t!]
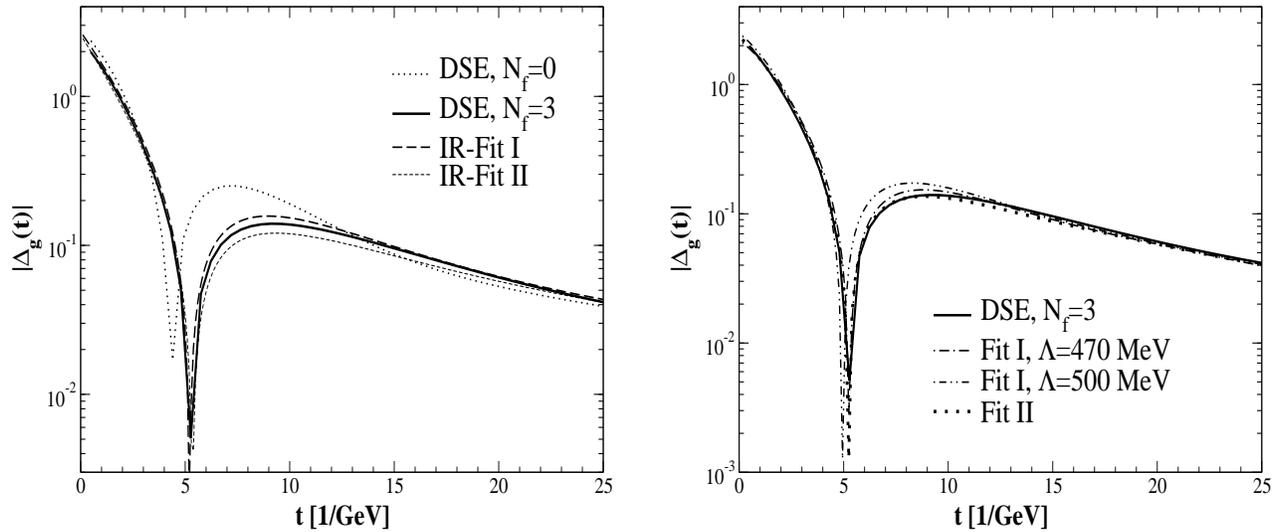

  \vspace{0.5cm} \centerline{
    \epsfig{file=p.gluon_pos3.eps,width=8cm,height=7cm}
    \hspace{0.5cm}
    \epsfig{file=p.gluon_pos4.eps,width=8cm,height=7cm} }
\caption{\label{gluon-pos.dat}
  The results for the absolute value of the gluon Schwinger function,
  $\Delta_g(t)$, corresponding to our numerical results from the DSEs
  are shown and compared to the fits in the infrared (left panel) and
  the overall fits (right panel).  The spikes mark the time scales where
  the Schwinger functions cross zero and negative norm contributions
  appear.}
\end{figure}
%%%%%%%%%%%%%%%%%%%%%%%%%%%%%%%%%%%%%%%%%%%%%%%%%%%%%%%%%%%%%%%%%%%%%%%%%%

Employing a numerical Fourier transform routine, we can now calculate
the Schwinger function, $\Delta_g(t)$ (defined by
Eq.~(\ref{schwinger})), for the numerical solutions of the gluon DSE
and for the various fits.  The absolute values of the numerical
Schwinger functions for $N_f=0,\,3$ (using transverse projection
$\zeta=1$) are displayed in Fig.~\ref{gluon-pos.dat}.  The spikes mark
the time scales at which the Schwinger functions cross zero and
negative norm contributions appear in each gluon propagator.  One
notes that the Schwinger function in the quenched approximation
differs visibly from that for three flavors, despite the similarity of
the corresponding gluon dressing functions for Euclidean momenta
\cite{Fischer:2003rp}.  In particular, the typical time scale, marked
by the zero of the Schwinger function, decreases from $5.2\;{\rm
  GeV}^{-1}$ to $4.4\;{\rm GeV}^{-1}$.  We have also explicitly
checked that different choices for the projection of the gluon
equation and other minor details of the truncation scheme lead only to
minor quantitative alterations (see Appendix A).  All gluon Schwinger
functions we have calculated from the results of the coupled DSEs show
the same qualitative behavior, thus demonstrating that neither the
details of the projection in the gluon equation nor the feedback of (a
small number of) dynamical quarks\footnote{The infrared ($p^2\to0^+$)
  behavior of the Yang--Mills sector of QCD is unaffected by the
  appearance of chiral quarks as long as the number of flavors is
  small enough to be in the confining and chiral symmetry breaking
  phase of QCD \cite{Fischer:2003rp}.}  have any significant influence
on the overall analytic structure of the gluon propagator. {\em We
  clearly observe positivity violations in the gluon propagator.}
This is the first major result of this work.

\subsection{Analytic structure of the gluon propagator}

In the following we aim at an interpretation of our results in terms
of the analytic structure of the gluon propagator in the timelike
momentum region.  As a first step we demonstrate that the infrared
behavior of the gluon propagator, {\em i.e.\/} the behavior for
$p^2\to 0^+$, is responsible for the non-trivial analytic structure.
To this end, in the left-hand side of Fig.~\ref{gluon-pos.dat} the
numerical results for the gluon Schwinger function (with
$\zeta=1,\,N_f=3$) are compared to the infrared fits,
Eqs.~(\ref{IRfitI}) and (\ref{IRfitII}).  The fitted parameters are
$w_I = 2.5$, $\Lambda_I=400$ MeV for IR-fit I, Eq.~(\ref{IRfitI}), and
$w_{II} = 2.7$, $\Lambda_{II}=420$ MeV for IR-fit II,
Eq.~(\ref{IRfitII}).  As we observed earlier, the fits only agree with
the numerical gluon dressing function in the infrared momentum region.
Nevertheless, in Fig.~\ref{gluon-pos.dat} we see that the agreement of
the numerical Schwinger function with the Fourier transforms of each
of these fits is excellent.  It appears that the details of the
intermediate and large momentum behavior of the gluon propagator have
little or no influence on the qualitative analytical structure of the
propagator in the ``near-by'' timelike momentum regime.  In
particular, the change in curvature at the bump of the gluon dressing
function at a scale of $\sim1$~GeV is not an important feature in this
regard.  In fact the crucial property of the gluon propagator is that
it goes to zero for vanishing momentum.  This can be seen easily as
the relation,
\begin{equation} 
  0 = D(p=0) = \int {d^4x} \; D(x) \,,
\end{equation}
(with $D(p) = Z(p^2)/p^2$) implies that the propagator function in
coordinate space, $D(x)$, must contain positive as well as negative
norm contributions, with equal integrated strengths.

For fit I (Eq.~(\ref{fullfit})) we have used two parameter sets, $w_I
= 2.4$, $\Lambda_{QCD}=500$ MeV and $w_I = 2.0$, $\Lambda_{QCD}=470$
MeV.  The first parameter set fits the gluon renormalization function
better (especially in the ultraviolet) and the second set is optimized
to fit the Schwinger function.  For fit II (Eq.~(\ref{fullfit})) with
the parameters $w_{II} = 2.5$ and $\Lambda_{QCD}=510$ MeV both the
gluon renormalization function and the Schwinger function are fitted
very well.  As the infrared fits I and II already reproduce the gluon
Schwinger function it is no surprise that the complete fits,
Eq.~(\ref{fullfit}), do even better, see the right-hand side of
Fig.~\ref{gluon-pos.dat}.  As already stated, for the sake of
simplicity we have used only one common scale, $\Lambda_{QCD}$, for
the infrared and ultraviolet behavior.

We are now in a position to deduce the possible analytic structure of
the gluon propagator.  We first observe that because of the infrared
singularity of the ghost propagator, we expect a cut on the timelike
momentum axis coming from the ghost-loop contribution to transverse
gluons.  As the ghost loop is the infrared dominant contribution in
the gluon equation and therefore determines the infrared behavior of
the gluon propagator, it is instructive to discuss the infrared fits
to the gluon propagator first.  The infrared fit I
(Eq.~(\ref{IRfitI})) contains a branch cut on the negative $p^2$ axis
while the denominator contributes a pair of complex conjugate
singularities at
\begin{equation} 
 p^2=\Lambda_{I}^2 \: e^{\pm i\pi/(2 \kappa)} \,.  
\end{equation}
The discontinuity across the negative $p^2$ axis is easily calculated.
Writing $p^2_\pm=(-\rho \pm i \epsilon)\Lambda_{I}^2$ one obtains
\begin{equation} 
 \lim_{\epsilon \rightarrow 0} 
  \left\{D_{I}(p^2_+)-D_{I}(p^2_-)\right\} \;=\; 
  \frac{-2i\omega_{I}}{\Lambda_I^2} \;
   \frac{\sin(2 \pi \kappa) \; \rho^{2\kappa-1}} 
        {1+2 \,\rho^{2\kappa} \,\cos(2 \pi \kappa) + \rho^{4 \kappa}} \,,
\end{equation}
with $D_{I}(p^2) = Z_{I}^{ir}(p^2)/p^2$.  This discontinuity rises from
zero at $\rho=0$ to a maximum at the area of the pole locations and
then rapidly decays as $\rho$ becomes larger.
  
In the infrared fit II (Eq.~(\ref{IRfitII})) the numerator and the
denominator conspire to produce one cut\footnote{Note that we have
decided to take the ratio first and then we raise it to a non-integer
power.  Having this non-integer for the numerator and the denominator
separately would lead to two overlapping branch cuts.  However, we
consider this an unnecessary complication.} over
$p^2\in(0,-\Lambda_{II}^2)$.  For the discontinuity we have (now for
$p^2_\pm=(-\rho \pm i \epsilon)\Lambda_{II}^2$)
\begin{equation} 
 \lim_{\epsilon \rightarrow 0} 
  \left\{D_{II}(p^2_+)-D_{II}(p^2_-)\right\} \;=\; 
  \frac{-2i\omega_{II}}{\Lambda_{II}^2} \;
   \frac{\sin(2 \pi \kappa) \; \rho^{2\kappa-1}}{(1-\rho)^{2\kappa}}
        \,, 
\end{equation} 
with $D_{II}(p^2) = Z_{II}^{ir}(p^2)/p^2$ and for $0 < \rho < 1$ only.
This rapidly diverges as $p^2 \downarrow -\Lambda_{II}^2$
(i.e. $\rho\rightarrow 1$) and then drops discontinuously to zero:
there is no discontinuity for $p^2 < -\Lambda_{II}^2$.

Whereas the location of the singularity $p^2=-\Lambda_{II}^2$ in the
infrared fit II is independent of the value of the exponent $\kappa$,
the location of the complex conjugate singularities of IR-fit I as
well as the magnitudes of the cuts in both fits depend on $\kappa$ and
therefore on the truncation scheme.  Although the exact value of
$\kappa$ depends on the details of the truncation, various methods
suggest that the exponent $\kappa$ is in the range $0.5 < \kappa < 1$
\cite{Lerche:2002ep,Zwanziger:2001kw,Fischer:2002hn,Pawlowski:2003hq}.
It is exactly this range which corresponds to the pair of complex
conjugate singularities in IR-fit I being located on the first Riemann
sheet in the left half of the complex $p^2$-plane.  In the limiting
case $\kappa=0.5$, one obtains one real pole on the negative
$p^2$-axis in both fits, and in the other limit, $\kappa=1$, IR-fit I
corresponds to a pair of purely imaginary poles, {\em i.e.\/} exactly
the form proposed in
Refs.~\cite{Gribov:1978wm,Zwanziger:mf,Stingl:1996nk}.

To discuss the analytic structure of the full fits,
Eq.~(\ref{fullfit}), we must also look at the analytic properties of
the expression for the running coupling, Eq.~(\ref{fitB}).  The Landau
pole at spacelike $p^2=\Lambda_{QCD}^2$ has been subtracted, so
expression (\ref{fitB}) only has singularities on the timelike real
axis.  The logarithm produces a cut on this half-axis, and the
corresponding discontinuity vanishes for $p^2\to 0^-$, diverges at
$p^2=-\Lambda_{QCD}^2$ and goes to zero for $p^2\to \infty$.  In the
fits I and II, Eq.~(\ref{fullfit}), the running coupling
(Eq.~(\ref{fitB})) is raised to a non-integer power and multiplied by
the infrared fits (Eqs.~(\ref{IRfitI}) and (\ref{IRfitII})).  Thus,
fit I also has a pair of complex conjugate singularities, at the same
locations as those in Eq.~(\ref{IRfitI}).  On the other hand, fit II
has no non-analyticities other than the cut on the negative real axis.
The discontinuity corresponding to the cut of the combination of the
different factors in fit II is always positive, vanishes for $p^2\to
0^-$, diverges at $p^2=-\Lambda_{QCD}^2$ to $+\infty$ and falls to
zero for $p^2\to -\infty$.

It is interesting to note the scale at which positivity violations
occur.  From Fig.~\ref{gluon-pos.dat} we determine that the zero
crossing appears at $t \approx 5\; \mbox{GeV}^{-1} \approx 1\;
\mbox{fm}$.  This is roughly the size of a hadron and therefore the
correct scale at which gluon screening should occur.  One might
speculate whether this represents an inherent, gauge invariant scale
(as the locations of propagator poles are protected by Nielsen
identities \cite{Nielsen:fs}), which is generated in the
renormalization process.  The pure power law $Z(p^2)=(p^2)^{2\kappa}$,
which solves the system of DSEs in the case where the renormalization
point $\mu$ is shifted to asymptotic values, is in perfect agreement
with the scale-invariance of the underlying theory, corresponding to
an infinite mass gap.  Thus it is obvious that we can deduce the
existence of a cut from the pure power laws, but we can not extract
the related scale.  This scale emerges from an interplay of infrared
and ultraviolet properties of the theory, {\it i.e.}  the transition
of the gluon propagator from the infrared power law to its
perturbative ultraviolet behavior.

Before concluding this subsection we comment on what lattice
Monte--Carlo simulations say about positivity violation in the gauge
boson propagator.  For unquenched QCD, nothing is known because the
gluon propagator has not yet been calculated with dynamical fermions.
The pure Yang--Mills gauge propagator has been calculated on the
lattice for almost twenty years following the pioneering work of
Mandula and Ogilvie \cite{Mandula:rh}, see {\em e.g.\/}
Refs.~\cite{Langfeld:2002dd,Furui:2003jr,Bonnet:2001uh,lattice} and
references therein.  However, explicit observations of positivity
violation have been elusive as statistical errors and finite volume
artefacts cloud the issue.  Nevertheless, many hints of negative norm
contributions in the gluon propagator have been reviewed in
\cite{Mandula:nj}.  Clear measurements of positivity violation have
been made for the case of $SU(2)$ \cite{Langfeld:2001cz} and for the
gluon propagator in three-dimensional Yang-Mills theory \cite{CucchieriPC}.

Summarizing: the Landau gauge gluon propagator, as it results from the
solution of coupled DSEs, displays positivity violations.  This is in
accordance with gluons being confined.  The infrared behavior of the
gluon propagator is analytically determined to be a power law.  It has
been demonstrated in Ref.~\cite{Lerche:2002ep} that this behavior is
stable under a broad range of possible dressings of the ghost-gluon
vertex.  Furthermore, strong arguments have been presented in
Ref.~\cite{Zwanziger:2003cf} for the existence of power laws in
generalized truncations that include the four-gluon interaction.  The
power law behavior at small Euclidean momenta induces a cut on the
real negative $p^2$-axis, as can be seen clearly from our infrared
fits.  It is this cut which causes the observed pattern of positivity
violation.  Fitting the gluon propagator for all Euclidean momenta
{\em and} the corresponding Schwinger function we are able to describe
the gluon propagator with fit II, Eq.~(\ref{fullfit}), which has no
singularities in the complex $p^2$-plane except for a cut on the
negative real axis.

Note that this fit contains essentially two parameters: the overall
magnitude which, because of renormalization properties, is
arbitrary,\footnote{{\em i.e.\/} it is determined via the choice of
the renormalization scale $\mu$ and the normalization condition
$G^2(\mu^2,\mu^2)Z(\mu^2,\mu^2)=1$.} and the scale $\Lambda_{QCD}$.
The infrared exponent, $\kappa$, and the anomalous dimension of the
gluon, $\gamma$, are not free parameters: $\kappa$ is determined from
the infrared properties of the DSEs and the one-loop value is used for
$\gamma$.  Therefore, we have found a parameterization of the gluon
propagator which has effectively only one physical parameter, the
scale $\Lambda_{QCD}$.  Combined with the relatively simple analytic
structure of fit II, Eq.~(\ref{fullfit}), this gives us confidence
that we have succeeded in uncovering the most important features of
the Landau gauge gluon propagator.

\subsection{Results for the quark propagator}

\begin{figure}[t!]
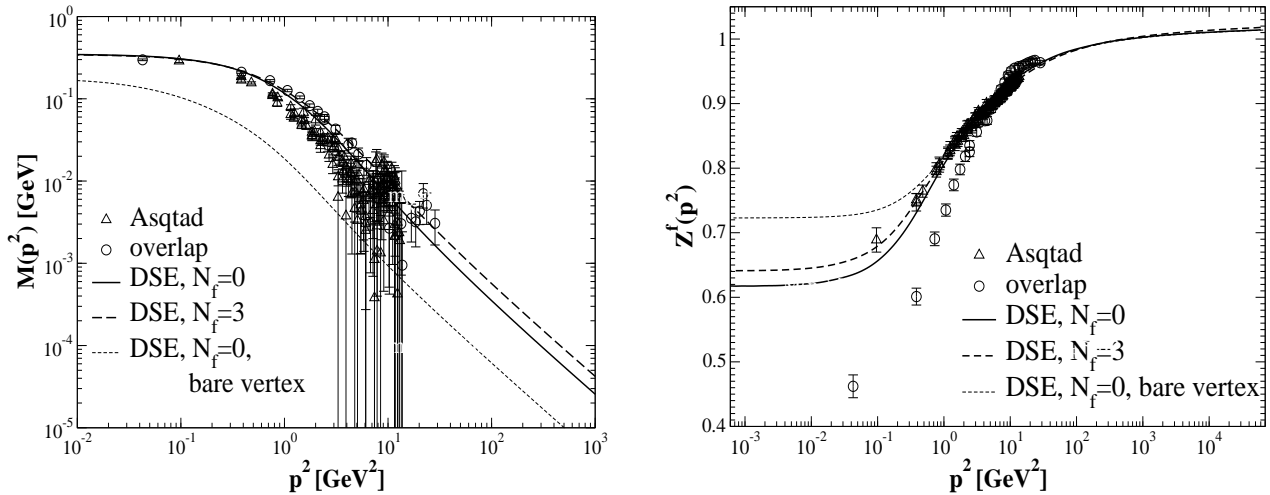

  \vspace{5mm} \centerline{
    \epsfig{file=p.quark.M.eps,width=80mm,height=65mm}
    \hspace{5mm}
    \epsfig{file=p.quark.A.eps,width=80mm,height=65mm} }
\caption{\label{quark.dat}
  The quark mass function, $M(p^2)$, and the wave function
  renormalization, $Z^f(p^2)$, from quenched $(N_f=0)$ and unquenched
  ($N_f=3$ chiral quarks) DSEs \cite{Fischer:2003rp}. Results for the
  generalized CP vertex, Eq.~(\ref{vertex_CP}), and the bare vertex
  construction, Eq.~(\ref{vertex_bare}), are compared with quenched
  lattice data in the overlap \cite{Bonnet:2002ih} and Asqtad
  \cite{Bowman:2002bm} formulations . }
\end{figure}

In Fig.~\ref{quark.dat} we display the mass function,
$M(p^2)=B(p^2)/A(p^2)$, and the wave function renormalization,
$Z^f(p^2)=1/A(p^2)$ (note the superscript $f$ which differentiates
this function from the gluon dressing function), of the quark
propagator in the chiral limit, obtained from the coupled quark,
ghost, and gluon DSEs \cite{Fischer:2003rp}.  We show quenched
$(N_f=0)$ and unquenched $(N_f=3)$ results employing the generalized
CP vertex, Eqs.~(\ref{vertex-ansatz})-(\ref{vertex_CP}).  We also
display the same functions calculated in the quenched approximation
with the bare Abelian part of the quark gluon vertex,
Eq.~(\ref{vertex_bare}).  On the Euclidean real axis, both vertex
constructions lead to qualitatively similar but quantitatively quite
different results.  The bare vertex approximation does not give enough
chiral symmetry breaking and is clearly disfavored by recent quenched
lattice data \cite{Bonnet:2002ih,Bowman:2002bm} (also shown in
Fig.~\ref{quark.dat}).  On the other hand, the results for the more
elaborate vertex construction are well within the region suggested by
the lattice calculations.

%%%%%%%%%%%%%%%%%%%%%%%%%%%%%%%%%%%%%%%%%%%%%%%%%%%%%%%%%%%%%%%%%%%%%%%%%%%%%%
%%%%%%%%%%%%%%%%%%%%%%%%%%%%%%%%%%%%%%%%%%%%%%%%%%%%%%%%%%%%%%%%%%%%%%%%%%%%

\begin{figure}[t!]
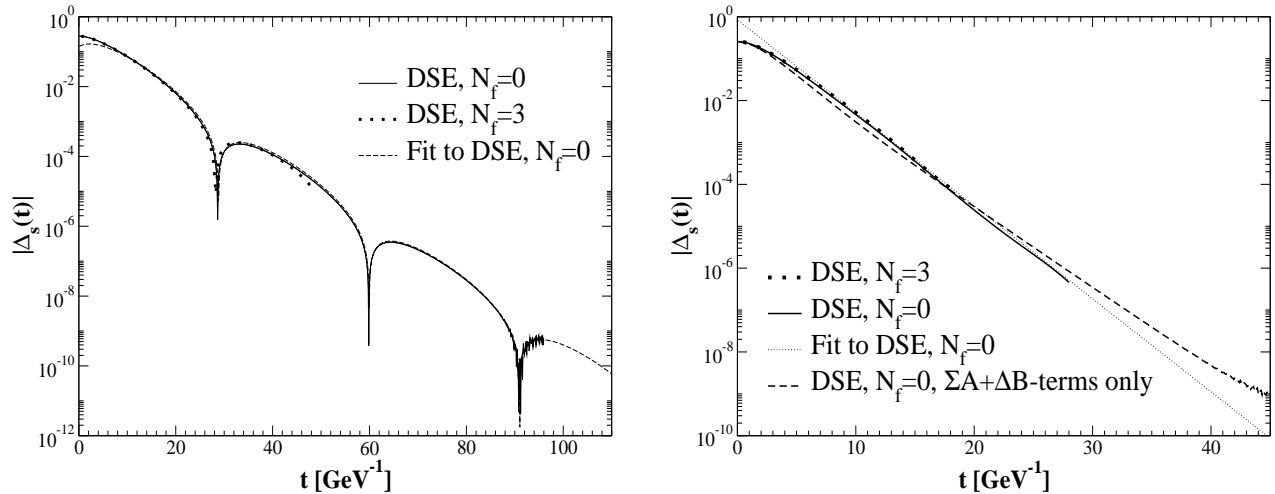

  \vspace{5mm} \centerline{
    \epsfig{file=p.bare.sigmaS_pos.eps,width=80mm,height=65mm}
    \hspace{5mm}
    \epsfig{file=p.cp.sigmaS_pos.eps,width=80mm,height=65mm} }
\caption{\label{quark_pos.dat}
  The left diagram displays the absolute value of $\Delta_s(t)$ 
  employing the bare vertex construction in the quark DSE.  
  The spikes correspond to zero crossings of the Schwinger function.  
  These are absent in the diagram on the right where the results 
  with the full CP vertex, Eq.~(\protect{\ref{vertex-ansatz}}), 
  are considered.  The chiral limit results are shown for $N_f=3$ 
  and $N_f=0$, together with the fits to the Schwinger function of 
  the quenched DSE solution. 
  Furthermore we compare to a calculation with only the two most 
  important terms of the quark-gluon vertex.}
\end{figure}

The quantitative difference between the DSE solutions using the bare
vertex and the CP vertex turns into a qualitative difference for the
corresponding Schwinger functions.  The Fourier transformed scalar
parts of the different quark propagators, $\Delta_s(t)$, are shown in
Fig.~\ref{quark_pos.dat}.  Similar results are obtained for the vector
parts of the propagators, $\Delta_v(t)$, though they are numerically
less accurate.\footnote{In the chiral limit, the scalar part of the
propagator, $\sigma_s(p^2)$, falls off like $1/p^4$, up to logarithmic
corrections, because the function $B(p^2)$ falls off like $1/p^2$,
whereas $\sigma_v(p^2)$ falls off like $1/p^2$.  This makes the
Fourier transform of the scalar part easier to calculate
numerically.}  As in the case of the gluon propagator, we plot the
absolute values of the Schwinger functions on a logarithmic scale. The
results in the left diagram are obtained employing the bare Abelian
part of the vertex, Eq.~(\ref{vertex_bare}).  Clearly these solutions
exhibit the oscillatory behavior of Eq.~(\ref{oszillation}), which is
characteristic for a propagator with a pair of complex conjugate
``mass-like'' singularities.  A fit of the expression in
Eq.~(\ref{oszillation}) to our result gives the locations of these
singularities as $\msing=(209 + 101 i)\;{\rm MeV}$.

A completely different picture is obtained from the Schwinger
functions constructed using the CP vertex, Eq.~(\ref{vertex_CP}), as
can be seen in the diagram on the right of Fig.~\ref{quark_pos.dat}.
Again we display results for the quenched case, $N_f=0$, and the case
of $N_f=3$ chiral quarks.  For $N_f = 0$ we also make use of a fit to
the running coupling as described in detail in
Ref.~\cite{Fischer:2003rp}; for all practical purposes the results are
almost indistinguishable from those obtained with the numerical
$\alpha(q^2)$ as a solution of the ghost-gluon DSEs.  We find no
traces of negative norm contributions, and in all cases, a fit of the
oscillatory form of Eq.~(\ref{oszillation}) to our results indicates
that there is a singularity (almost) on the real timelike axis, with
an imaginary part of at most $8\%$ of its real part.  The best fit is
obtained for a real mass singularity at $\msing=0.50\;{\rm GeV}$.  For
both the bare and the CP vertex, the deviation of the fits from the
data at small time scales suggests that there is additional structure
in the DSE solution which the simple pole fits (Eqs.~(\ref{decay}) and
(\ref{oszillation})) do not capture.  We shall investigate this in
Sec.~\ref{sec:quarkparam}.

%%%%%%%%%%%%%%%%%%%%%%%%%%%%%%%%%%%%%%%%%%%%%%%%%%%%%%%%%%%%%%%%%%%%%%%%%%%%%
%%%%%%%%%%%%%%%%%%%%%%%%%%%%%%%%%%%%%%%%%%%%%%%%%%%%%%%%%%%%%%%%%%%%%%%%%%%%%%

By turning the different contributions in the vertex construction of
Eq.~(\ref{vertex_CP}) on and off, we have identified the term which is
responsible for the qualitative differences between the left and right
diagram of Fig.~\ref{quark_pos.dat}.  In addition to the (dominant)
vector part of the vertex
\begin{equation} 
  \Sigma A_\mu := \frac{A(p^2) + A(q^2)}{2} \gamma_\mu\,, 
\label{SumA}
\end{equation}
the presence of the scalar coupling $\Delta B_\mu$,
Eq.~(\ref{DeltaB}), in the quark-gluon vertex is crucial for the
substantial change in the analytic structure of the quark propagator
compared to the truncation keeping only the vector part.  Such a
scalar term introduces additional feedback in the scalar self-energy,
and its presence considerably enhances the amount of dynamical chiral
symmetry breaking generated in the quark DSE.  By varying the strength
of this term compared to the leading $\Sigma A_\mu$-piece of the
vertex, we find that a reduction of this term by about $20\%$ is
enough to generate again positivity violations corresponding to
dominant complex conjugate singularities.

\begin{table}
\begin{ruledtabular}
\begin{tabular}{l|ccc}
      & bare vertex  &  $\Sigma A_\mu + \Delta B_\mu$-term  &   CP vertex 
\\ \hline
YM $\alpha(k^2)$, unquenched, $N_f=3$ 
      & 0.21(1) $\pm$ 0.10(1) $i$   &  0.48(3)          & 0.50(3)
\\
YM $\alpha(k^2)$, quenched ($N_f=0$) 
      & 0.21(1) $\pm$ 0.10(1) $i$   &  0.48(3)    & 0.50(3)
\\
fit A of Ref.~\cite{Fischer:2003rp}, quenched
      & 0.209(4) $\pm$ 0.101(2) $i$ &  0.48(3)          & 0.50(3)
\\
fit B of Ref.~\cite{Fischer:2003rp}, quenched
      & 0.160(4) $\pm$ 0.076(2) $i$ &  0.42(3)          & 0.42(3)
\\
Maris--Tandy model~\cite{Maris:1999nt}, Eq.~(\ref{model1}) 
      & 0.55(1) $\pm$ 0.321(6) $i$  & 0.96(6)   & 1.1(1)
\\
Gaussian model~\cite{Alkofer:2002bp}, Eq.~(\ref{model2}) 
      & 0.53(1) $\pm$ 0.167(3) $i$  & 0.83(4)   & 0.83(6) 
\\
quenched QED (in units of $10^{-3}\Lambda$)
      & 1.79(6) $\pm$ 0.43(2) $i$   &           & 1.51(9) 
\\
\end{tabular}
\end{ruledtabular}
\caption{\label{mass-table} 
  Results for the fermion pole masses in the chiral limit for different 
  interactions, as extracted from the behavior of the corresponding 
  Schwinger functions.  The quark masses are given in GeV,  the 
  QED$_4$ results are given in units of the UV cutoff $\Lambda$,
  and are obtained with $\bar\alpha=1.2$ for the bare vertex
  and $\bar\alpha=1.06$ for the CP vertex.  The errors are estimates
  of the total numerical error; the numerical error in case
  of a real mass singularity is dominated by the fact that, on 
  a logarithmic scale, the Schwinger functions are not perfect
  straight lines.}
\end{table}

The question of positivity violation does not depend on the details of
the input from the Yang--Mills sector of QCD.  We obtain
quantitatively similar results for the unquenched case with $N_f=3$
chiral quarks, for the quenched approximation with the running
coupling taken directly from the Yang--Mills DSEs and for different
models for the running coupling \cite{Fischer:2003rp}.\footnote{We
have even arbitrarily changed $\alpha(0)$ from its value 2.97 in these
fits.  Dynamical chiral symmetry breaking occurs for $\alpha(0)>
\alpha_{\hbox{\scriptsize crit}}$ with $\alpha_{\hbox{\scriptsize
crit}}$ being slightly below one. For $\alpha(0)$ in the range
$\alpha_{\hbox{\scriptsize crit}}< \alpha(0) < 10$ we found no
evidence for positivity violation when the CP vertex is used.}  As a
check, we also employ the model interactions given in
Eqs.~(\ref{model1}) and (\ref{model2}).  Again we obtain evidence for
a pair of complex conjugate singularities when a bare vertex is used
and a singularity on the real timelike momentum axis once the
additional scalar coupling is taken into account.\footnote{Note that a
similar result has been found in the model study of
Ref.~\cite{Hawes:ef} where a Stingl-type gluon propagator model has
been employed in the quark DSE together with a quark-gluon vertex
consisting only of the Abelian Ball--Chiu and Curtis--Pennington type
structures~\cite{Curtis:1990zs,Ball:ay}.  In this study the absence of
complex singularities in the quark propagator has been attributed to
the vanishing of the employed model gluon propagator at zero momentum.
This interpretation seemed to be supported by a study using the same
propagator and a bare vertex which finds also real
poles~\cite{Bender:1994bv}.  However, the present study clearly
demonstrates that for a sufficiently strong interaction the crucial
reason for this absence of complex singularities lies in the
quark-gluon vertex.}  Our results for the pole masses obtained in
these models are given in Table~\ref{mass-table}.  For the model
interaction Eq.~(\ref{model1}) we agree with the estimate for the
singularity closest to $p^2=0$ given in Ref.~\cite{Jarecke:2002xd}
based on a Taylor series expansion of the quark propagator functions,
confirming that we can indeed extract the location of the first
singularity via the Schwinger functions.  Finally, we checked the
truncation scheme of Ref.~\cite{Bloch:2002eq} where a model
interaction with an infrared finite coupling has been employed
together with a bare quark-gluon vertex.  In this case we also found a
pair of complex conjugate poles as could be expected.

\begin{figure}
  \vspace{5mm}\centerline{
    \epsfig{file=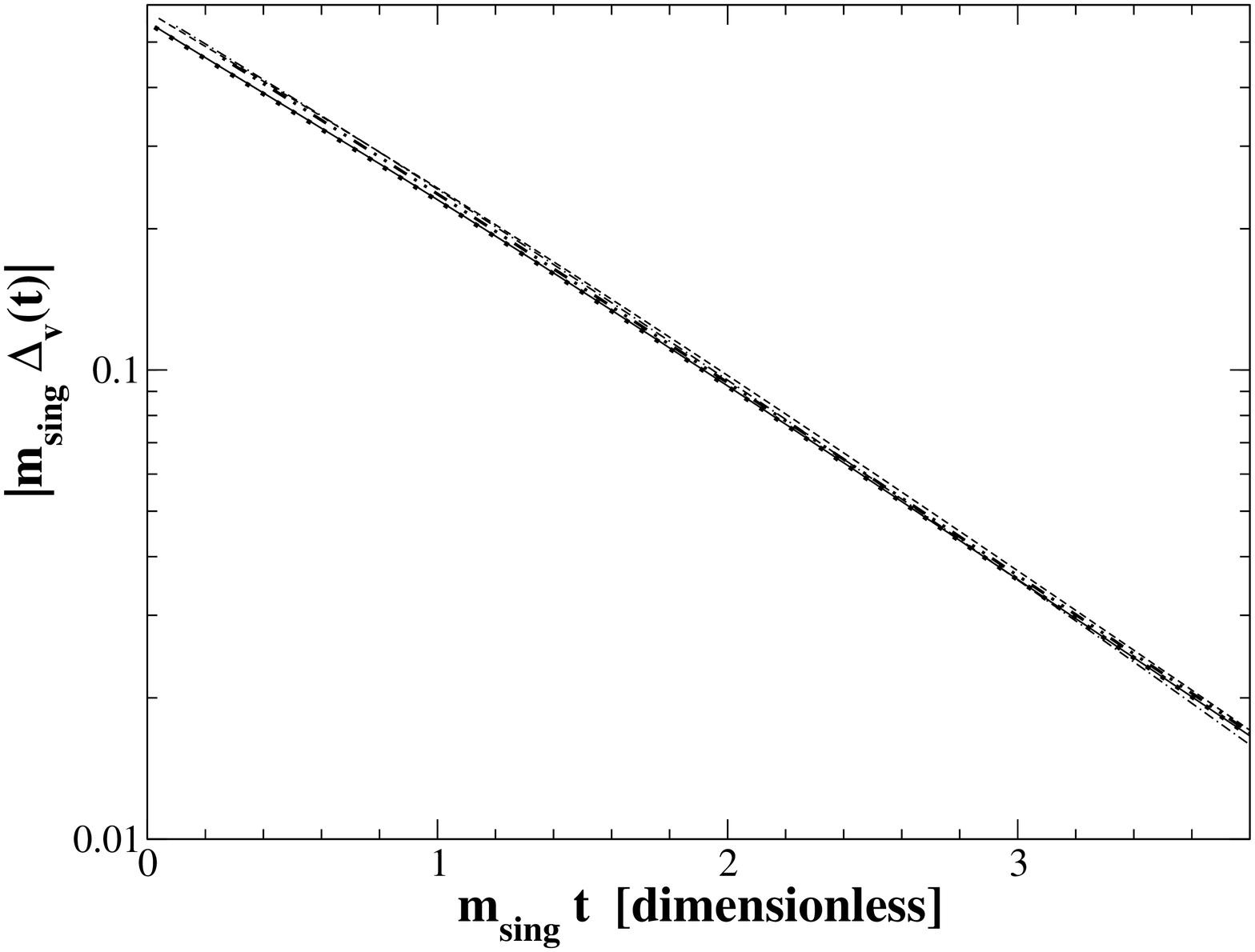,width=80mm,height=65mm}
    \hspace{5mm}
    \epsfig{file=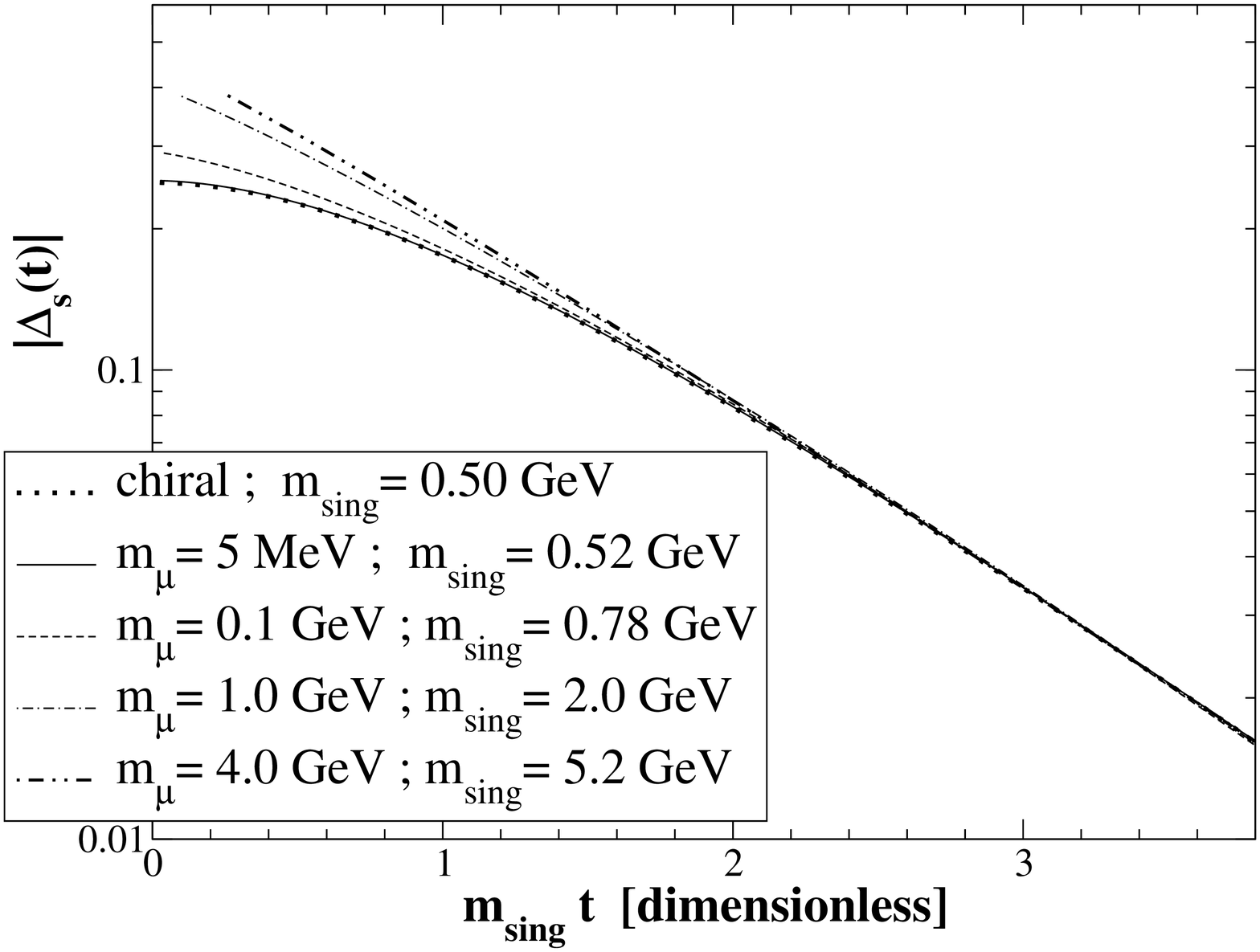,width=80mm,height=65mm}
  }
  \caption{\label{fig:masspos}
    The dimensionless Schwinger functions $\msing \Delta_v(t)$ (left) and 
    $\Delta_s(t)$ (right) as function of $\msing\,t$, where $\msing$ is 
    the ``pole mass'' as determined by the exponential decay of the 
    Schwinger function for different current quark masses $m_\mu$, 
    renormalized at $\mu = 10\;{\rm GeV}$. }
\end{figure}

Another interesting property of expression (\ref{DeltaB}) is its
insensitivity to explicit chiral symmetry breaking, {\em i.e.\/} a
current quark mass.  The contributions from current quark masses to
the function $B(p^2)$ are almost momentum independent and therefore
cancel quite accurately in Eq.~(\ref{DeltaB}).  The Schwinger
functions become steeper with increasing quark mass, but show no signs
of positivity violation, even for current quark masses as large as a
few GeV.  For a detailed comparison of the mass dependence of the
Schwinger functions $\Delta_s(t)$ and $\Delta_v(t)$, we scale
$\Delta_v(t)$ by the pole mass, $\msing$ (extracted from the
exponential decay of $\Delta_{s,v}(t)$), and plot $\Delta_s(t)$ and
$\msing \Delta_v(t)$ as function of the dimensionless variable
$\msing\,t$ in Fig.~\ref{fig:masspos}.  This reveals that the only
mass dependence is in the curvature of $\Delta_s(t)$ at small
$\msing\,t$: with increasing current quark mass the amount of
curvature decreases.

%%%%%%%%%%%%%%%%%%%%%%%%%%%%%%%%%%%%%%%%%%%%%%%%%%%%%%%%%%%%%%%%%%%%%%
%%%%%%%%%%%%%%%%%%%%%%%%%%%%%%%%%%%%%%%%%%%%%%%%%%%%%%%%%%%%%%%%%%%%%%

How can we understand this curvature that is present in $\Delta_s(t)$
but not in $\Delta_v(t)$?  A possible origin could be the fact that
the function $\sigma_s(p^2)$ drops off like $1/q^4$ in the chiral
limit while $\sigma_v(q^2)$ decreases as $1/q^2$.  As can be seen from
Eq.~(\ref{decay}), a single real pole on the negative momentum axis
results in a pure, exponential decay of the corresponding Schwinger
function.  However, the Schwinger function of a propagator with two
poles is
\begin{equation}
 \frac{1}{\pi}\int_0^\infty
  dp \cos(t p) \frac{1}{p^2 + m^2} \; \frac{1}{p^2 + \Lambda^2} \; = \;
     \frac{1}{2(\Lambda^2 - m^2)}
     \left(\frac{1}{m} e^{-mt}  - \frac{1}{\Lambda} e^{-\Lambda t} \right)
     \,,
\end{equation}
and for $\Lambda$ somewhat larger than $m$, this could lead to the
observed curvature at small $t$.  This, in combination with the fact
that this curvature tends to decrease with increasing current quark
mass, suggests that this curvature is related to the $1/p^2$ fall off
(up to logarithmic corrections) of $M(p^2)$ in the chiral limit.
However, there are other mechanisms that could generate such curvature
as we will discuss in more detail in the next section.

Comparing the two panels of Fig.~\ref{fig:masspos}, we also see that
$\Delta_s(t)$ approaches $\msing \Delta_v(t)$ from below for all
values of the current quark mass.  In other words, we find that
(within numerical accuracy) $\msing\Delta_v(t) > \Delta_s(t)$ for all
$t$.  Based on the constraint for the spectral decomposition,
Eq.~(\ref{eq:spectralvsconstraint}), this is what one would expect for
a propagator describing a Dirac field with asymptotic states.  Thus,
within this approach there are no signals of positivity violation in
the non-perturbative quark propagator.

Considering these findings, we state the second major result of this
work: {\em the presence of a scalar quark-gluon coupling of sufficient
  strength leads to a positive definite quark propagator with a
  singularity on the timelike real momentum axis.}  As our quark-gluon
vertex has been constructed as an {\em ansatz}, we do not have model
independent information on the relative strength of the different
tensor structures in the true quark-gluon vertex.  Our assumption has
been that all non-Abelian corrections can be accounted for by an
overall factor multiplying an Abelian construction for the tensor
structure of the vertex (see Eq.~(\ref{vertex-ansatz})).  This
factorization assumption has been tested in a recent investigation of
the quark-gluon vertex in quenched lattice QCD and was found to be
only valid at a qualitative level~\cite{Skullerud:2003qu}.  However,
as yet no definite statements can be extracted from the lattice
calculations as they are only performed in two special kinematical
situations, whereas in our calculations the vertex is probed over the
whole range of momenta.  Further investigations are necessary to
determine the relative strength of the various components of the
vertex in a model independent manner.

\begin{figure}[t!]
  \vspace{5mm} \centerline{
    \epsfig{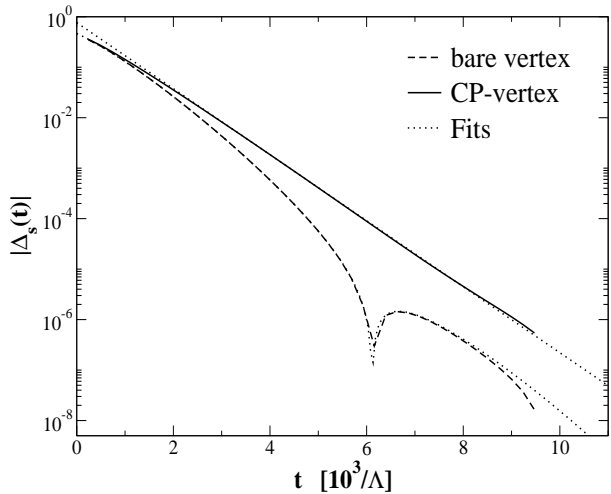}
  }
\caption{\label{qed.dat}
  Results for two different vertex constructions in QED$_4$. For ease
  of comparison we employed two different values for the coupling,
  {\em i.e.\/} $\bar\alpha=1.2$ in the case of the bare vertex and
  $\bar\alpha=1.06$ for the case of the CP vertex.}
\end{figure}

In QED$_4$ however, we encounter a somewhat different situation.  The
vertex construction is more constrained than in QCD as the
longitudinal part of the CP vertex, the Ball--Chiu
vertex~\cite{Ball:ay}, is exact and the relative strengths of the
three longitudinal Dirac structures in the vertex are uniquely
determined by the Ward identity, Eq.~(\ref{QED-quark-gluon-STI}).  The
results for the fermion propagator in quenched approximation
($\alpha(q^2)\equiv\overline{\alpha}$, constant) in the chirally
broken phase of quenched QED are very similar to those of QCD.  Again,
we find a fermion propagator that satisfies positivity as long as it
is calculated with a vertex obtained from the Ward identity but
violates positivity if a bare vertex is used.  The Schwinger functions
are shown in Fig.~\ref{qed.dat} and the deduced (complex) pole masses
are included in Table~\ref{mass-table}.  Of course, it remains
possible that the transverse parts of the exact vertex conspire to
lead to positivity violation again.  However, this is unlikely, in
particular in QED where one has no confinement.

%%%%%%%%%%%%%%%%%%%%%%%%%%%%%%%%%%%%%%%%%%%%%%%%%%%%%%%%%%%%%%%%%%
%%%%%%%%%%%%%%%%%%%%%%%%%%%%%%%%%%%%%%%%%%%%%%%%%%%%%%%%%%%%%%%%%%%%%%%%

\section{Analytic properties of the quark propagator from
parameterizations \label{sec:quarkparam} }

In this section we explore the possible analytic structure of the
quark propagator in more detail.  Here we also consider the available
lattice data for the quark propagator and investigate whether it is
possible to obtain information on the analytic structure of the
propagator by fitting this data, the DSE solutions, and the
corresponding Schwinger functions with different parameterizations of
pole locations and/or branch cuts.  The singularity on the real
momentum axis may be accompanied by additional real singularities at
larger mass scaled or by complex conjugate singularities with a
larger real part of the mass, or it may be the starting point of a
branch cut on the negative real momentum axis.  In the next two
subsections we explore these possibilities.

\subsection{Meromorphic parameterizations}

The most rigorous constraint on the non-perturbative quark propagator
is that it must reduce to a free fermion propagator at large momenta
because of asymptotic freedom.  This entails that the propagator
functions,
$\sigma_{s,v}(p^2)\stackrel{|p^2|\to\infty}{\longrightarrow}0$ in {\em
all} directions of the complex $p^2$-plane \cite{Oehme:1996ju}.
Additionally, the theory of complex functions tells us that if
$\sigma_v(p^2)$ and $\sigma_s(p^2)$ are not constant, they cannot be
analytic over the whole complex plane: non-constant, entire functions
which are analytic at all finite points in the complex plane are
already excluded by the asymptotic properties of the propagator
functions.  From the truncated set of DSEs explored in the previous
section, we found the dominant (in terms of the Schwinger function)
structure to be either a singularity on the negative real $p^2$ axis
or a pair of complex conjugate singularities in the left half of the
complex $p^2$-plane.  In both scenarios the poles are accompanied by
additional undetermined structures which are responsible for the small
time behavior of $\Delta_{s}(t)$.  Guided by these results we first
consider parameterizations of the renormalized quark propagator using
the meromorphic form
\begin{equation}
  S(p)=Z_2^{-1}\sum_{j=1}^{n_P}
  \left(\frac{r_j}{i\pslash +a_j +i b_j} +
    \frac{r_j}{i\pslash +a_j -i b_j} \right)\,, 
  \label{cc-form}
\end{equation}
with $n_P$ pairs of complex conjugate poles located at $a_j\pm ib_j$
with residues $r_j$.  This form includes the possibility of complex
conjugate as well as purely real poles, but enforces neither of these
from the outset.  Similar simple parameterizations have been considered
in Refs.~\cite{Bhagwat:2003wu}.

In the following, we use physical constraints as well as lattice data
to fix the position of the various singularities.  The only practical
restriction on this procedure is in the number of parameters that can
be pinned down.  As further simplifications, we assume that the
residues, $r_j$, of these poles are real (although this is not a
strict requirement) and only consider the chiral limit.

For the propagator functions, $\sigma_s(p^2)$ and $\sigma_v(p^2)$, the
form Eq.~(\ref{cc-form}) simplifies to
\begin{eqnarray}        
  \sigma_v(p^2) &=& Z_2^{-1}
  \sum_{j=1}^{n_P}\frac{2r_j(p^2+a_j^2-b_j^2)}
  {(p^2+a_j^2-b_j^2)^2 +4a_j^2b_j^2}\,, 
  \label{ParV}\\
  \sigma_s(p^2) &=& Z_2^{-1}
  \sum_{j=1}^{n_P}\frac{2r_ja_j(p^2+a_j^2+b_j^2)}
  {(p^2+a_j^2-b_j^2)^2 +4a_j^2b_j^2}\,.
  \label{ParS}
\end{eqnarray}  
In terms of these quantities, we can construct the usual
renormalization point independent mass function
$M(p^2)=\sigma_s(p^2)/\sigma_v(p^2)$ and the wave-function
renormalization $Z^{f}(p^2)=(p^2+M^2(p^2))\sigma_v(p^2)$.  In order to
make contact with lattice data (where the finite lattice spacing leads
to a maximum possible momentum), we renormalize at $\mu^2=16$~GeV$^2$.

There are various restrictions we can impose on the parameters $r_j$,
$a_j$ and $b_j$ in the meromorphic form, Eq.~(\ref{cc-form}).  These
arise from its mathematical properties, from experimental observables
and from recent lattice data.  Asymptotic freedom requires that quarks
behave like free particles at large momenta.  Consideration of the
large momentum limit of $\sigma_v(p^2)$ implies that
\begin{equation}
        \sum_{j=1}^{n_P}r_j=\frac{1}{2}\,. \label{as1}
\end{equation}
Since we are working in the chiral limit, the mass function, $M(p^2)$,
must vanish for large spacelike real momenta. This entails that\footnote{
If we move away from the chiral limit, the right hand side of
Eq.~(\ref{as2}) is replaced by the renormalized current mass.}
\begin{equation}
        \sum_{j=1}^{n_P}r_ja_j=0\,. \label{as2}
\end{equation}
Furthermore, $M(p^2\to+\infty)$ must be real and approach zero from above. 

Asymptotically, the chiral limit mass function behaves as
\cite{CONDENSATE}
\begin{equation}
  M(p^2)\stackrel{p^2\to\infty}{\longrightarrow}\frac{2\pi^2\gamma_m}{N_c}
  \frac{-\langle\bar{q}q\rangle}{p^2\left[\frac{1}{2}
      \ln(\frac{p^2}{\Lambda^2_{QCD}})\right]^{1-\gamma_m}}\,,
\label{qqasymp} 
\end{equation}
where $\langle\bar{q}q\rangle$ is the renormalization-point-invariant
chiral condensate.  Although the logarithmic behavior of
Eq.~(\ref{qqasymp}) cannot be reproduced by these simple meromorphic
fits, the logarithm is a slowly varying function and we estimate the
condensate by fitting the mass function with Eq.~(\ref{qqasymp}) over
the range $p^2\in(10^3, 10^9)$ using the $\Lambda_{QCD}=0.5\;{\rm
GeV}$ and the appropriate 1-loop value of $\gamma_m=12/33$ for $N_f =
0$.  We then insist that this condensate extracted from our
meromorphic propagator agrees with the phenomenological value
$\langle\bar{q}q\rangle=-[0.275(75)\,{\rm GeV}]^3$.

In order to be phenomenologically applicable, the propagator should
reproduce the pion decay constant to a reasonable accuracy.  To
calculate this, we employ the approximation~\cite{Roberts:dr},
\begin{equation} 
  f_\pi^2 \simeq Z_2 \frac{N_c}{4\pi^2}\int_0^{\Lambda^2} dp^2 p^2
  \frac{M(p^2)}{Z^f(p^2)}\left[\sigma_v(p^2)\sigma_s(p^2)
    +\frac{p^2}{2}\left(\frac{d\sigma_v(p^2)}{dp^2}\sigma_s(p^2)-
      \sigma_v(p^2)\frac{d\sigma_s(p^2)}{dp^2}\right)\right]\,,
  \label{fpi}
\end{equation}
which incorporates only the effects of the leading Dirac structure of
the pion Bethe--Salpeter amplitude in the chiral limit.  From a
comparison of the relative sizes of the pion Bethe--Salpeter
amplitudes in model calculations~\cite{Maris:1997tm,Alkofer:2002bp},
one concludes that this approximation should lead to an
underestimation of $f_\pi$ by 10-20 \%.\footnote{One also knows from
chiral perturbation theory that the chiral limit pion decay constant
is somewhat less than the physical value of 93~MeV.}  In our
meromorphic fits we therefore demand that Eq.~(\ref{fpi}) gives
$f_\pi\sim 0.08(3)$~GeV.

The Landau gauge quark propagator has been investigated on the lattice
by a number of different groups using mean-field- and
non-perturbatively- improved clover actions \cite{Skullerud:2000un},
the Kogut--Susskind action \cite{Bowman:2002bm}, the overlap formalism
\cite{Bonnet:2002ih} and the Asqtad quark action \cite{Bowman:2002bm}.
The data sets obtained in the latter two formulations have the
smallest error bars and are therefore employed in what follows.  Their
mass functions and wave-function renormalizations have already been
shown in Fig.~\ref{quark.dat}.  The mass function data from the
lattice have been quadratically extrapolated
\cite{Bonnet:2002ih,Bowman:2002bm} to the chiral limit, whereas the
mass dependence of $Z^f(p^2)$ is very mild so no extrapolation has
been performed.  While the simple extrapolation procedure that has
been employed may lead to sizable errors \cite{Bhagwat:2003vw}, it
will prove sufficient for our purposes.

Unfortunately all of the lattice studies to date make use of the
quenched approximation.  Removing all internal quark loops is a
potentially drastic modification of the theory.  It destroys the
unitarity of the S-matrix, however it is often assumed that these
violations of unitarity are small.  Strictly speaking, it is
nonsensical to discuss the concept of positivity in such a situation
and the lattice data discussed above cannot be relied on to provide
any guidance in studying positivity of the quark propagator.  However,
from our experiences with the DSE studies of the previous section, one
may expect that quenching will not qualitatively change the momentum
dependence of the propagator (see Fig.~\ref{quark.dat}).
Additionally, the lattice data apparently still contain large finite
volume effects (especially in the wave-function renormalization)
\cite{Bowman:2002kn}, and do not precisely constrain the asymptotic
($p^2\to+ \infty$) behavior of the propagator .  For these reasons we
do not directly fit the lattice data (though {\em a posteriori}
$\chi^2$ fits to it return very similar parameters to those we find
below), but merely extract its three qualitative infrared features.
Thus we assume that the zero momentum values of the mass function and
wave-function renormalization, $M_0$ and $Z^f_0$, and an approximate
width of the region of large dynamical mass generation, $\omega_L$
(defined by $M(\omega_L^2)=M_0/2$), are robust against the effects of
quenching (within substantial errors).  With this in mind, we require
that our parametric fits are in reasonable agreement with the
extracted values of $M_0$, $Z^f_0$ and $\omega_L$.  That is:
\begin{equation}
  \label{lattvals}
  M_0=0.35(10) \:\mbox{GeV}\,, \quad\quad Z^f_0=0.6(2)\,, 
  \quad\quad \omega_L=0.7(2) \:\mbox{GeV}\,.
\end{equation}
Note that $\langle\bar{q}q\rangle$, $f_\pi$ and these three parameters
are obviously not entirely unrelated.

Given the number of independent constraints we can impose, we can
reasonably expect to be able to determine only five or six parameters.
This implies $n_P\le3$ in Eq.~(\ref{cc-form}).  We find that three
paradigmatic cases satisfy the requirements of
Eqs.~(\ref{as1})--(\ref{lattvals}): three purely real poles (denoted,
3R), two pairs of complex poles (2CC), and a real pole plus a pair of
complex conjugate poles (1R+1CC).  In order to construct the best fits
for each of these forms, we first impose the simple constraints of
Eqs.~(\ref{as1}) and (\ref{as2}) to reduce the number of parameters to
be varied.  Then for each parameterization we randomly sample the
available parameter space, constructing a large ensemble of parameter
sets that satisfy the full set of constraints.  The best fit
parameters and their errors are finally calculated as the mean and
standard deviation of the parameters in this ensemble.

The simplest possible parameterizations of a single real pole or a
single pair of complex conjugate poles ($n_P=1$ in
Eq.~(\ref{cc-form})) cannot satisfy the required constraints.
Specifically, enforcing the perturbative asymptotic behavior
(Eqs.~(\ref{as1}) and (\ref{as2})) makes it impossible to satisfy any
of the other requirements described above. Similarly, for two real
poles ($n_P=2$, $b_1=b_2=0$), the restrictions on the infrared
properties ($f_\pi$, $M_0$ and $\omega_L$) are incompatible with a
realistic quark condensate.

\begin{table}
\begin{ruledtabular}
\begin{tabular}{l|ccc|ccc|ccc}
& $r_1$ & $a_1$ [GeV]& $b_1$ [GeV]& $r_2$ & $a_2$ [GeV]& $b_2$ [GeV]
& $r_3$ & $a_3$ [GeV] \\  \hline
3R  
& 0.365(15) & 0.341(25) & --      & 1.2(8)     & -1.31(12) & --
& -1.06(*) & -1.40(*) & \\
2CC 
& 0.360(22) & 0.351(69) & 0.08(5) & 0.140(*)   & -0.899(*) & 0.463(75)
& -- & -- & \\
1R+1CC 
& 0.354(15) & 0.377(64) & --      & 0.146(*)   & -0.91(*) & 0.45(7) 
& -- & -- & \\
\end{tabular}
\end{ruledtabular}
\caption{\label{partab} Best fit parameters of the three meromorphic
  forms: three real poles (3R), two pairs of complex conjugate poles
  (2CC) and one real pole and one pair of complex complex conjugate
  poles (1R+1CC). The parameters whose errors are replaced by an
  asterisk, are completely determined in terms of the other parameters
  through Eqs.~(\ref{as1}) and (\ref{as2}).  In order to reproduce the
  results presented here, one should use the values that follow from
  Eqs.~(\ref{as1}) and (\ref{as2}) for those constrained parameters.}
\end{table}

\begin{table}
\begin{ruledtabular}
\begin{tabular}{l|ccccc}
& $M_0$ [GeV] & $Z^f_0$ & $\omega_L$ [GeV] & $f_\pi$ [GeV] &
$-\langle\bar{q}q\rangle^{1/3}$ [GeV] \\ \hline
3R     & 0.29(1) & 0.55(7) & 0.79(4) & 0.071(3) & 0.3(2) \\
2CC    & 0.33(11) & 0.57(12) & 0.69(27) & 0.070(31) & 0.3(3) \\
1R+1CC & 0.31(7) & 0.52(7) & 0.72(25) & 0.068(23) & 0.3(2) \\
\end{tabular}
\end{ruledtabular}
\caption{\label{restab} Values for the various constrained quantities
  for the three parameterizations of Table~\ref{partab}. Errors are
  solely due to uncertainties in the parameterizations and do not
  include any additional systematic errors.}
\end{table}

As mentioned above, a satisfactory realization of the requirements of
Eqs.~(\ref{as1})--(\ref{lattvals}) is possible in the case of three
real poles ($n_P=3$ and $b_1=b_2=b_3=0$).  The best fit parameters we
obtain are shown in Table~\ref{partab} and related quantities that
they result in are given in Table~\ref{restab}.  Although the
propagator functions $\sigma_{s,v}(p^2)$ have poles at
$p^2\sim-0.2\;\mbox{GeV}^2$, they exactly cancel in the combinations
$M(p^2)$ and $Z^f(p^2)$.  However the functions $M(p^2)$ and
$Z^f(p^2)$ do have poles further in the timelike region, the first one
occurring at $p^2\sim-0.75\;\mbox{GeV}^2$.  Also the zeros of
$Z^f(p^2)$ on the real axis may be problematic as they will
necessarily produce singularities in the CP construction of the
quark-gluon vertex, {\em c.f.\/} Eq.~(\ref{vertex_CP}).

In the case of two pairs of complex conjugate poles ($n_P=2$), the
best fit parameters and calculated quantities are again given in
Tables~\ref{partab} and \ref{restab}.  Both $M(p^2)$ and $Z^f(p^2)$
exhibit unexpected behavior around $p^2\sim-0.12 \;\mbox{GeV}^2$,
where they have a very sharp pole and a zero on the real axis. This
arises because $\sigma_s(p^2)$ and $\sigma_v(p^2)$ have zeros at very
slightly differing momenta ($p^2=-0.127\;\mbox{GeV}^2$ vs
$-0.117\;\mbox{GeV}^2$) and it may be somewhat troublesome.  This
behavior, as well as the small imaginary part of the location of the
first pair of poles, suggests forcing the first pair of poles to
collapse to one real pole ($n_P=2$, $b_1=0$).

\begin{figure}
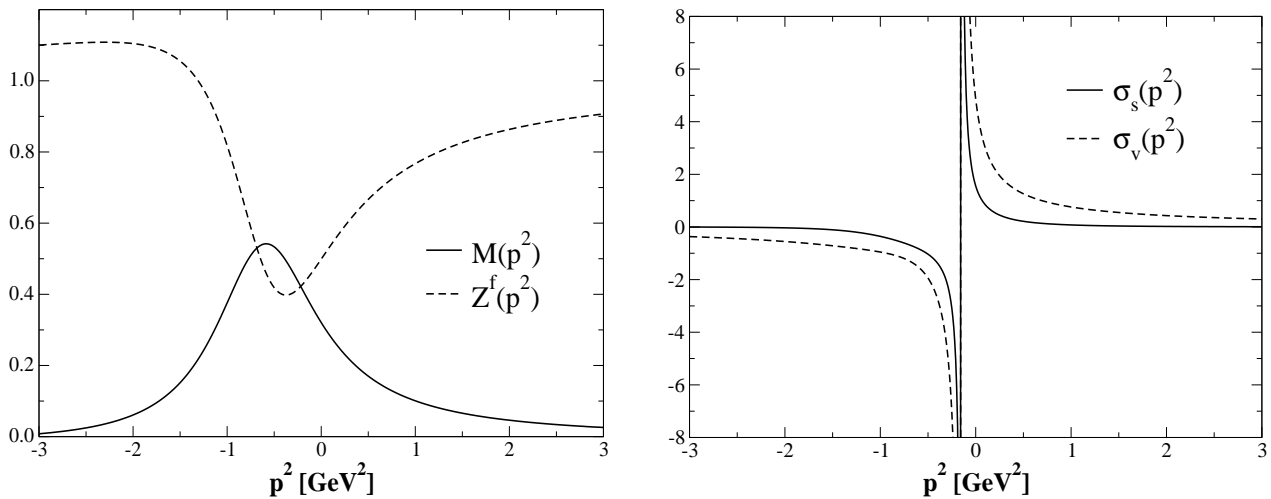

  \vspace{5mm} \centerline{
    \epsfig{file=polfit.rc.mz.eps,width=80mm,height=65mm}
    \hspace{5mm}
    \epsfig{file=polfit.rc.sv.eps,width=80mm,height=65mm} }
\caption{\label{rcpoles} Propagator functions for the fit using one real
  pole and one pair of complex conjugate poles.}
\end{figure}

Redoing the fits with one real pole and one pair of complex conjugate
poles, we come up with very similar parameters to the 2CC
parameterization, as listed in Table~\ref{partab}.  The corresponding
propagator functions are shown in Fig.~\ref{rcpoles}.  With this
parameterization, the strange behavior of $M(p^2)$ and $Z^f(p^2)$
disappears and $Z^f(p^2)$ only has complex conjugate poles and zeros
($Z^f(p^2=-0.41\pm0.48i\;\mbox{GeV}^2)=0$,
$Z^f(p^2=-0.55\pm0.69i\;\mbox{GeV}^2)\to\infty$) so the longitudinal
part of the quark-gluon vertex, Eq.~(\ref{vertex_CP}) will not have
particle-like singularities \cite{Roberts:2000hi}.  This
parameterization also contains one parameter less than the others.
Therefore we consider this to be the preferred form of the meromorphic
parameterizations investigated here.

%%%%%%%%%%%%%%%%%%%%%%%%%%%%%%%%%%%%%%%%%%%%%%%%%%%%%%%%%%%%%%%%%%%%%%%%%%%
%%%%%%%%%%%%%%%%%%%%%%%%%%%%%%%%%%%%%%%%%%%%%%%%%%%%%%%%%%%%%%%%%%%%%%%%%%%

\begin{figure}
  \vspace{5mm}\centerline{
    \epsfig{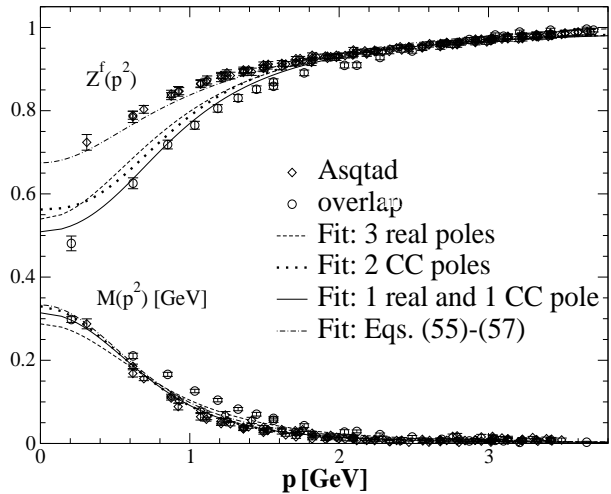} }
\caption{\label{latcomp} The best fit curves for the meromorphic
  parameterizations compared to the lattice data.  For the details
  of the parameterization of $\sigma_{s,v}(p^2)$ by a form with a branch
  cut, fitted to the Asqtad data, see the next subsection.}
\end{figure}

In comparing the three sets of parameterizations, it is worth
remarking that the location of the (real part of the) first pole and
its residue are extremely robust.  The obtained value for this
constituent quark mass, $m = 377(64)\;{\rm MeV}$ for our best fit, is
also in good agreement with a value extracted from lattice
simulations of the quark propagator using a tree-level Symanzik
improved action, $m = 342(13)\;{\rm MeV}$ \cite{Karsch:1998xd}.
However, the constraints on the other features in the fits are less
precise, especially in the case of three real poles.  In
Fig.~\ref{latcomp} we compare the parameterizations given in
Table~\ref{partab} to the lattice data; overall, the agreement is
quite acceptable.  Note that the meromorphic fits have relatively low
values of $Z^f_0$; this may change once finite volume effects are
reduced in the lattice data.  Also, each parameterization has a
somewhat low value of $f_\pi$ in the chiral limit.  This can be
attributed on the one hand to the approximation leading to
Eq.~(\ref{fpi}), and on the other hand to the approximations on the
lattice: the chiral extrapolation as well as the omission of dynamical
quarks might lead to an underestimation of $f_\pi$ in the lattice data
\cite{Bhagwat:2003vw}.

%%%%%%%%%%%%%%%%%%%%%%%%%%%%%%%%%%%%%%%%%%%%%%%%%%%%%%%%%%%%%%%%%%%%%%%%%%%

%%%%%%%%%%%%%%%%%%%%%%%%%%%%%%%%%%%%%%%%%%%%%%%%%%%%%%%%%%%%%%%%%%%%%%%%%%%

Having determined the best parameters for three different forms of our
fit functions, we now examine the Fourier transforms of the momentum
space propagator functions $\sigma_{s,v}(p^2)$.  Specifically, we
attempt to determine whether the sub-dominant behavior of the various
parameterizations can be determined from the Schwinger function, and,
if so, apply this to the DSE solutions of Sec.~\ref{sec:DSEsolns}.

Using the identity
\begin{equation}
  \int_0^\infty dx \frac{\cos(x y)}{x^2+c^2} = \frac{\pi}{2c}\exp(-c\,y) 
  \quad\quad\quad \left[y\in \mathbb{R},\,\arg(c^2)\neq\pi \right]\,,
\end{equation}
we can directly calculate the Schwinger functions from our
parameterizations (Eqs.~(\ref{ParV}) and (\ref{ParS})):
\begin{eqnarray}
  \Delta_s(t)&=& \sum_{i=1}^{n_P} sgn(a_i)r_ie^{-|a_i|t}
  \cos(b_it)\,,  
  \label{eq:fts}\\
  \Delta_v(t)&=& \sum_{i=1}^{n_P} \frac{r_ie^{-|a_i|t}}{a_i^2+b_i^2}
  \left(|a_i|\cos(b_it)-b_i\sin(b_it)\right)\,.
  \label{eq:ftv}
\end{eqnarray}
For all parameterizations, the term with the smallest mass parameter
$a_i$ will dominate for large $t$.
%
%%%%%%%%%%%%%%%%%%%%%%%%%%%%%%%%%%%%%%%%%%%%%%%%%%%%%%%%%%%%%%%%%%%%%%
\begin{figure}
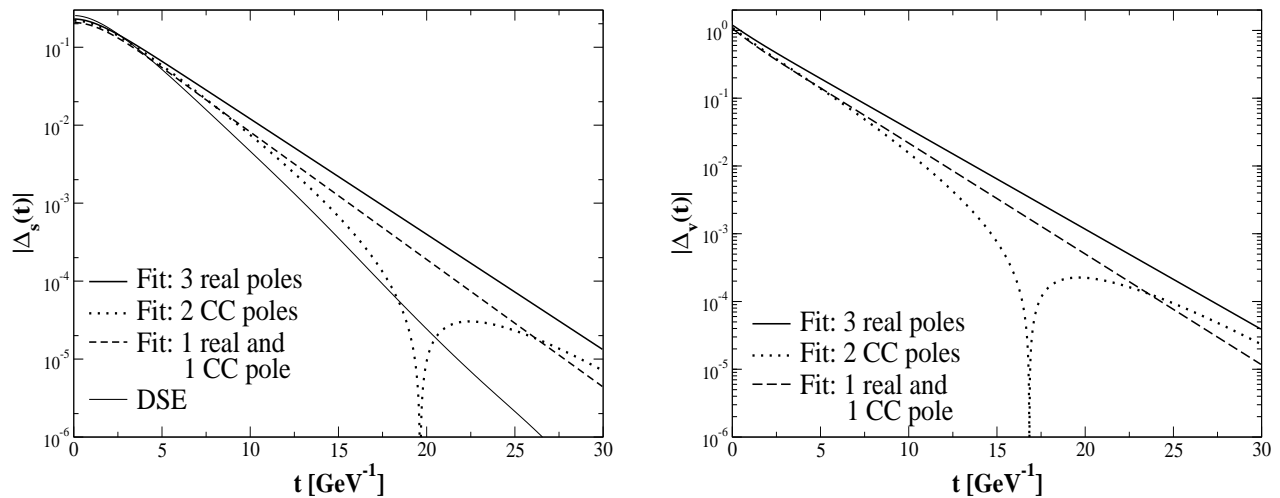

  \vspace{5mm}\centerline{
    \epsfig{file=p.polefitFFT.eps,width=80mm,height=65mm}
    \hspace{5mm}
    \epsfig{file=p.polefitFFT2.eps,width=80mm,height=65mm}
  }
  \caption{\label{fit.fft} The Fourier-transform of both $\sigma_s(p^2)$ 
    (left) and $\sigma_v(p^2)$ (right) for the optimal parameterizations
    with multiple poles.  In the diagram on the left hand side we 
    have included our DSE results.}
\end{figure}
%%%%%%%%%%%%%%%%%%%%%%%%%%%%%%%%%%%%%%%%%%%%%%%%%%%%%%%%%%%%%%%%%%%%%%
%
In Fig.~\ref{fit.fft} we display the analytic Fourier transforms of
the parameterized scalar and vector propagator functions,
Eqs.~(\ref{eq:fts}) and (\ref{eq:ftv}).  For comparison, we also
include our DSE result for $\Delta_s(t)$ employing the CP vertex.
Note the qualitative difference between the parameterization with two
complex conjugate poles and the other two.  Whereas the latter show no
sign of positivity violation, in the 2CC parameterization we clearly
see zero crossings of the Schwinger functions, both in $\Delta_s(t)$
and in $\Delta_v(t)$ (even a small imaginary component in the complex
conjugate masses is detectable provided the Fourier transform can be
calculated accurately to large enough $t$).  Note that $\Delta_s(t)$
calculated from the meromorphic parameterizations shows a similar
amount of small $t$ curvature to the DSE result, but $\Delta_v(t)$ is
linear in this region.  Thus multiple poles as explored here could
explain the small $t$ behavior observed in the DSE Schwinger
functions.

We also use these analytic Fourier transforms to test our numerical
Fourier transform, finding that it reproduces the analytic results
down to $\Delta_{s,v}(t) \sim 10^{-6}$ where we begin to run into
accuracy problems.  However, the numerical routine we employ is
clearly able to distinguish between a dominant real pole and dominant
complex conjugate poles.  This gives us further confidence that our
results from the DSE solutions in the previous section are not
numerical artefacts.

\subsection{Parameterizations with branch cuts}

As mentioned above, there is evidence that the Schwinger function
$\Delta_s(t)$ is convex (with sizable curvature) at small $t$.  On the
other hand, the Schwinger function $\Delta_v(t)$ as obtained from the
DSE solution shows no such curvature.  This difference could be
accommodated within the simple meromorphic fits of the previous
subsection.  However, this is certainly not the only possible
mechanism leading to such a difference, and here we explore the
consequences of allowing for singularities with branch cuts.  As can
be seen from Fig.~\ref{fig:masspos}, the curvature of $\Delta_s(t)$
depends on the current quark mass, so we also consider the effects of
explicit chiral symmetry breaking.

Our motivation for investigating such parameterizations arises from
considering the DSE for the quark propagator,
Eq.~(\ref{eq:truncDSEquark}).  If the combination
${\alpha(k^2)}/{k^2}$ is non-analytic at $k^2=(q-p)^2=0$ (in other
words, if $\alpha(0)\neq 0$), the integration path necessarily passes
through the external point $p$.  Thus, in order to evaluate the quark
propagator at arbitrary complex momenta, one has to deform the
integration contour in the DSE and solve the DSE along this deformed
integration path.  As long as there are no singularities in the other
factors of the integrand (i.e. in $S(q)$ and $V_\nu^{abel}(q,k)$),
this can in principle be done unambiguously (though it is numerically
a nontrivial task).  However, if we want to evaluate the integral for
a value $p$ at which the propagator, $S(p)$, has a singularity, we are
forced by the analytic structure of $\alpha(k^2)/{k^2}$ to include
this value of $p$ in the integration contour for $d^4q$. Thus, we have
a pinch singularity at this point coming from $S(q)$ and
$\alpha(k^2)/{k^2}$; this generally leads to a branch-cut, as is shown
in more detail in Appendix B.  We also note that the asymptotic form
of the quark propagator has perturbatively calculable logarithmic
contributions.  Considering these points, we would expect that the
singularities in $\sigma_{s,v}(p^2)$ are branch points rather than
simple poles.  Thus we next attempt to parameterize the quark
propagator by functions with branch cuts using the parameterization of
the strong running coupling, Eq.~(\ref{fitB}), that has proven helpful
in understanding the analytic structure of the gluon propagator.

As a first try, we shall fit the inverse propagator functions $M(p^2)$
and $Z^f(p^2)$ as obtained from the quark DSE with the CP vertex.
Given the close agreement of the DSE solutions and the lattice quark
propagator seen in Fig.~\ref{gluon.dat}, fitting the DSE solution
will result in similar physical constraints to those of the
previous subsection.  The leading-order perturbative behavior is
known, and we allow for one additional sub-leading term, that is to be
fitted to the DSE solution.  Furthermore, we want the parameterization
to have a branch cut along the negative real axis starting at $p^2 =
-\msing^2$.  Thus we are lead to fit the DSE solutions with
\begin{eqnarray}
Z^f(p^2) &=& Z_2 \left(1 - \frac{\alpha(p^2+\msing^2)}{2\pi} 
            + \frac{C_2}{p^2 + \msing^2 + \Lambda^2} \right) \,,
\label{eq:fitZ}
\\
M(p^2) &=& C_{dcsb} \;  \frac{\alpha(p^2 + \msing^2)^{1-\gamma_m}}
        {p^2 + \msing^2 + \Lambda^2}
    + \frac{C_4}{(p^2 + \msing^2 + \Lambda^2)^2}  
    + C_{cqm} \; \alpha(p^2 + \msing^2)^{\gamma_m} \,.
\label{eq:fitM} 
\end{eqnarray}
The parameters $C_{dcsb}$ and $C_{cqm}$ are related to the chiral
condensate and the renormalized current quark mass, respectively:
\begin{eqnarray}
-\langle\bar{q}q\rangle &=&  C_{dcsb} \; 
         \left(\frac{\pi\gamma_m}{2}\right)^{1-\gamma_m}\;
        \frac{N_c}{2\pi^2\gamma_m}  \,,
\\
 m_\mu &=& C_{cqm} \; \alpha(p^2+m^2)^{\gamma_m}  \;\; \approx \;\;
        C_{cqm} \; \left(\frac{\pi\gamma_m}
                        {\ln(p^2/\Lambda_{QCD}^2)}\right)^{\gamma_m} \,.
\end{eqnarray}
The renormalization constant $Z_2$ is determined by the
renormalization condition $Z^f(\mu^2)=1$, $\msing$ follows from the
exponential decay of the Schwinger functions, and we take $\Lambda$ to
be equal to $\Lambda_{QCD}$ in the running coupling, $\alpha(x)$, for
which we use Eq.~(\ref{fitB}).  The remaining free parameters in this
fit, $C_2$ and $C_4$, are fitted to the numerical solution of the DSE
and $\Lambda_{QCD}$ is also varied to improve this fit.

%%%%%%%%%%%%%%%%%%%%%%%%%%%%%%%%%%%%%%%%%%%%%%%%%%%%%%%%%%%%%%%%%%%%%%
\begin{figure}
  \vspace{5mm}\centerline{
    \epsfig{file=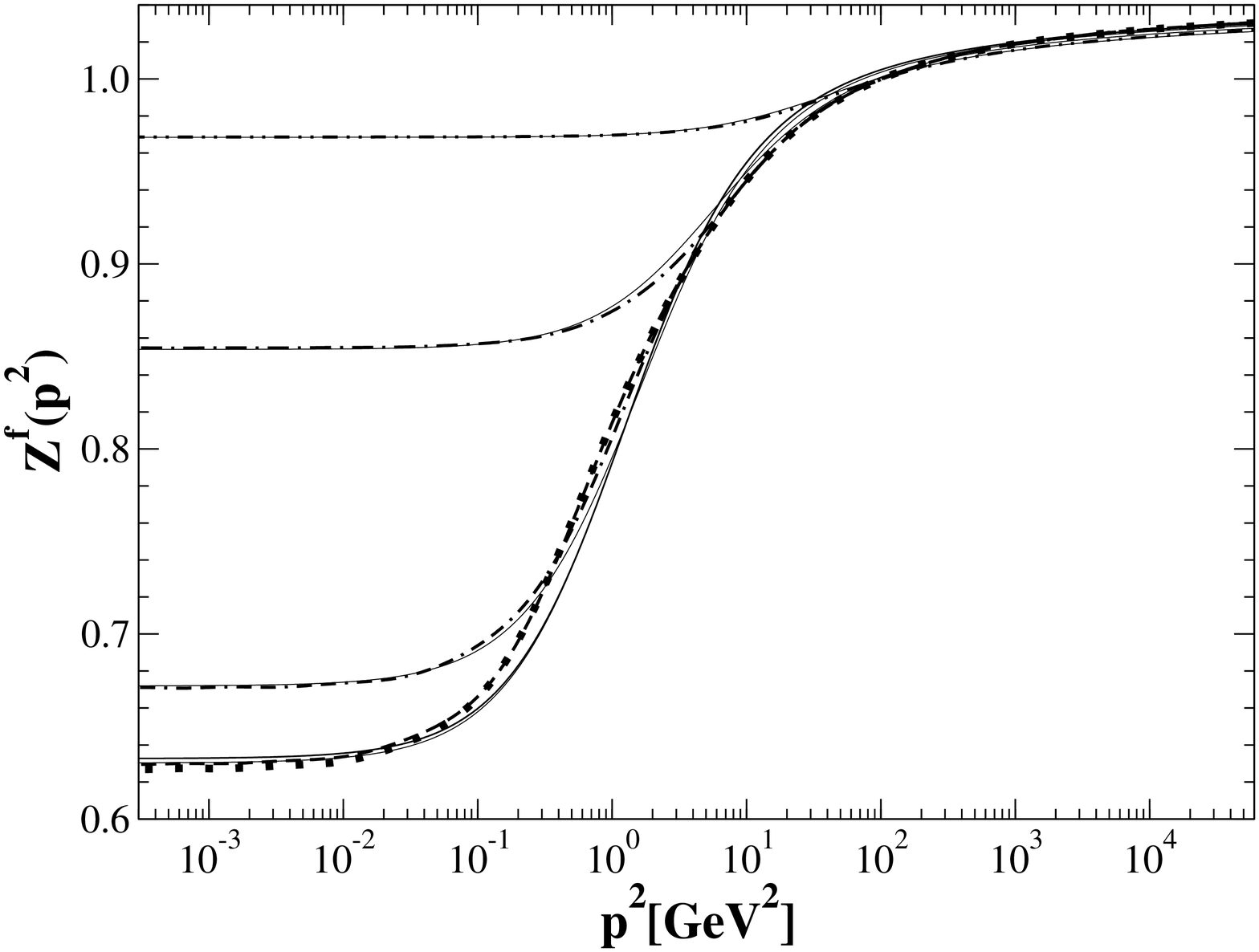,width=80mm,height=65mm}
    \hspace{5mm}
    \epsfig{file=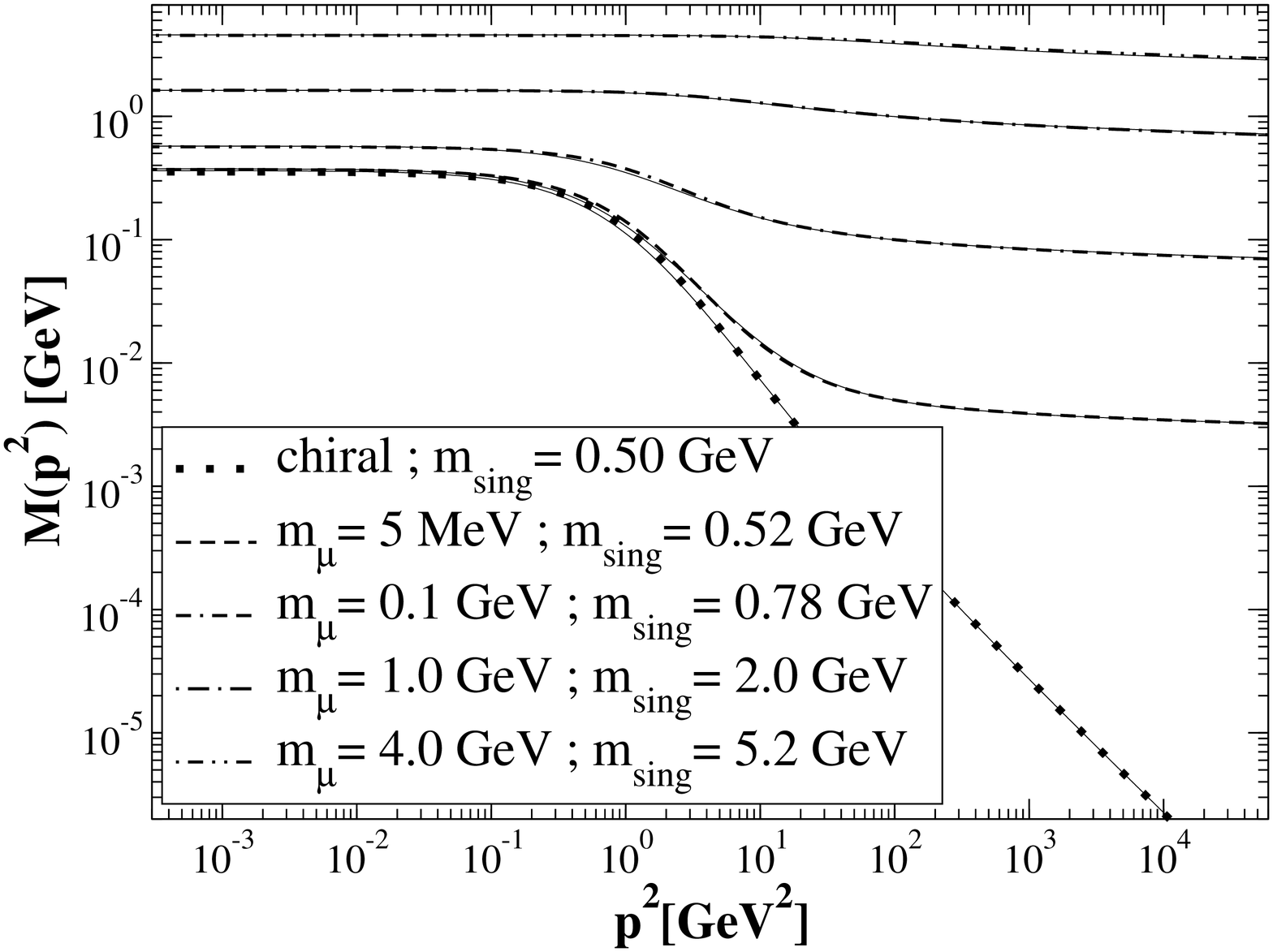,width=80mm,height=65mm}
  }
  \caption{\label{fig:DSEfitMZ}
    The functions $Z^f(p^2)$ (left) and $M(p^2)$ (right) obtained from 
    parameterizations fitted to the numerical DSE solutions for 
    different current quark masses.  The solid curves are the fits 
    (Eqs.~(\ref{eq:fitZ}) and (\ref{eq:fitM})) and the dotted, dashed, 
    and various dot-dashed curve are the DSE solutions as shown in the 
    legend (the curves for $Z^f(p^2)$ in the chiral limit and for 
    $m_\mu = 5\;{\rm MeV}$ are almost indistinguishable).  The fit 
    parameters are given in Table~\ref{table:parameters}.}
\end{figure}
%%%%%%%%%%%%%%%%%%%%%%%%%%%%%%%%%%%%%%%%%%%%%%%%%%%%%%%%%%%%%%%%%%%%%%
\begin{table}
\begin{ruledtabular}
\begin{tabular}{l|lllll}
 $m_\mu$ & $C_{dcsb}$ & $C_4$ & $C_{cqm}$ 
                 & $C_2$, $\tilde{C}$ & $\msing$ \\ \hline
\multicolumn{6}{l}{fitting $M$ and $Z^f$, 
Eqs.~(\ref{eq:fitZ}) and (\ref{eq:fitM}); 
$\Lambda_{QCD}^2 = 0.81\;{\rm GeV}^2$}
\\ \hline
DSE, chiral& 0.086 & 0.248 & 0    & -0.011 & 0.50 \\
  0.005   & 0.119 & 0.202 & 0.0074& -0.015 & 0.52 \\
  0.10    & 0.343 & 0     & 0.16  & -0.067 & 0.78 \\
  1.0     & 0.36  & 0     & 1.65  & -0.224 & 2.0 \\
  4.0     & 0     & 0     & 6.6   & -0.348 & 5.2  \\ \hline
\multicolumn{6}{l}{fitting $\sigma_{s,v}$, 
Eqs.~(\ref{eq:fitBchiral})--(\ref{eq:fitsigv}); 
$\Lambda_{QCD}^2 = 0.70\;{\rm GeV}^2$}  
\\ \hline
DSE, chiral& 0.086 & 0.234 & 0.0    & 1.27 & 0.50 \\
  0.005    & 0.10  & 0.234 & 0.0076 & 1.26 & 0.52 \\ 
  0.100    & 0.44  & 0     & 0.161  & 1.11 & 0.78 \\ \hline
\multicolumn{6}{l}{fitting $\sigma_{s,v}$, 
Eqs.~(\ref{eq:fitBchiral})--(\ref{eq:fitsigv}); 
$\Lambda_{QCD}^2 = 0.50\;{\rm GeV}^2$}
\\ \hline
lattice, chiral
           & 0.08  & 0.12  & 0.0    & 1.47 & 0.47 \\ \hline
\multicolumn{6}{l}{fitting $A$ and $B$, 
%restricted to $\msing= M(-\msing^2)$,
Eqs.~(\ref{eq:fitA})--(\ref{eq:mconstraint});
$\Lambda_{QCD}^2 = 0.70\;{\rm GeV}^2$}
\\ \hline
DSE, chiral& 0.09  & 0.31 & 0     & 0.25 & 0.49 \\
  0.005    & 0.10  & 0.30 & 0.008 & 0.26 & 0.50 \\
  0.100    & 0.33  & 0    & 0.17  & 0.25 & 0.65 \\ 
  1.0      & 0.21  & 0    & 1.7   & 0.23 & 1.74 \\ 
  4.0      & 0     & 0    & 6.7   & 0.34 & 5.1 \\ 
\end{tabular}
\end{ruledtabular}
\caption{\label{table:parameters} Parameters for the various branch
cut fits (Eqs. (\ref{eq:fitZ})--(\ref{eq:mconstraint})) to the $N_f=0$
quark DSE solutions for different masses.  A parameter set obtained by
fitting Eqs.~(\ref{eq:fitBchiral})--(\ref{eq:fitsigv}) to the Asqtad
lattice data \cite{Bowman:2002bm} is also included.}
\end{table}
%%%%%%%%%%%%%%%%%%%%%%%%%%%%%%%%%%%%%%%%%%%%%%%%%%%%%%%%%%%%%%%%%%%%%%

The results are shown in Fig.~\ref{fig:DSEfitMZ} for several different
current quark masses representative of masses up to that of the bottom
quark.  The fitted parameters are given in the first section of
Table~\ref{table:parameters}.  With only a few parameters, the fits
represent the DSE solutions very well over the entire Euclidean
region.  The fitted values of $C_{cqm}$ are all reasonably close to
the current quark masses that were used as input in the DSEs (small
deviations are due to sub-leading effects) and the (fitted) chiral
condensate is acceptable: $-\langle\bar{q}q\rangle = (290\;{\rm
MeV})^3$.

\begin{figure}
  \vspace{5mm}\centerline{
    \epsfig{file=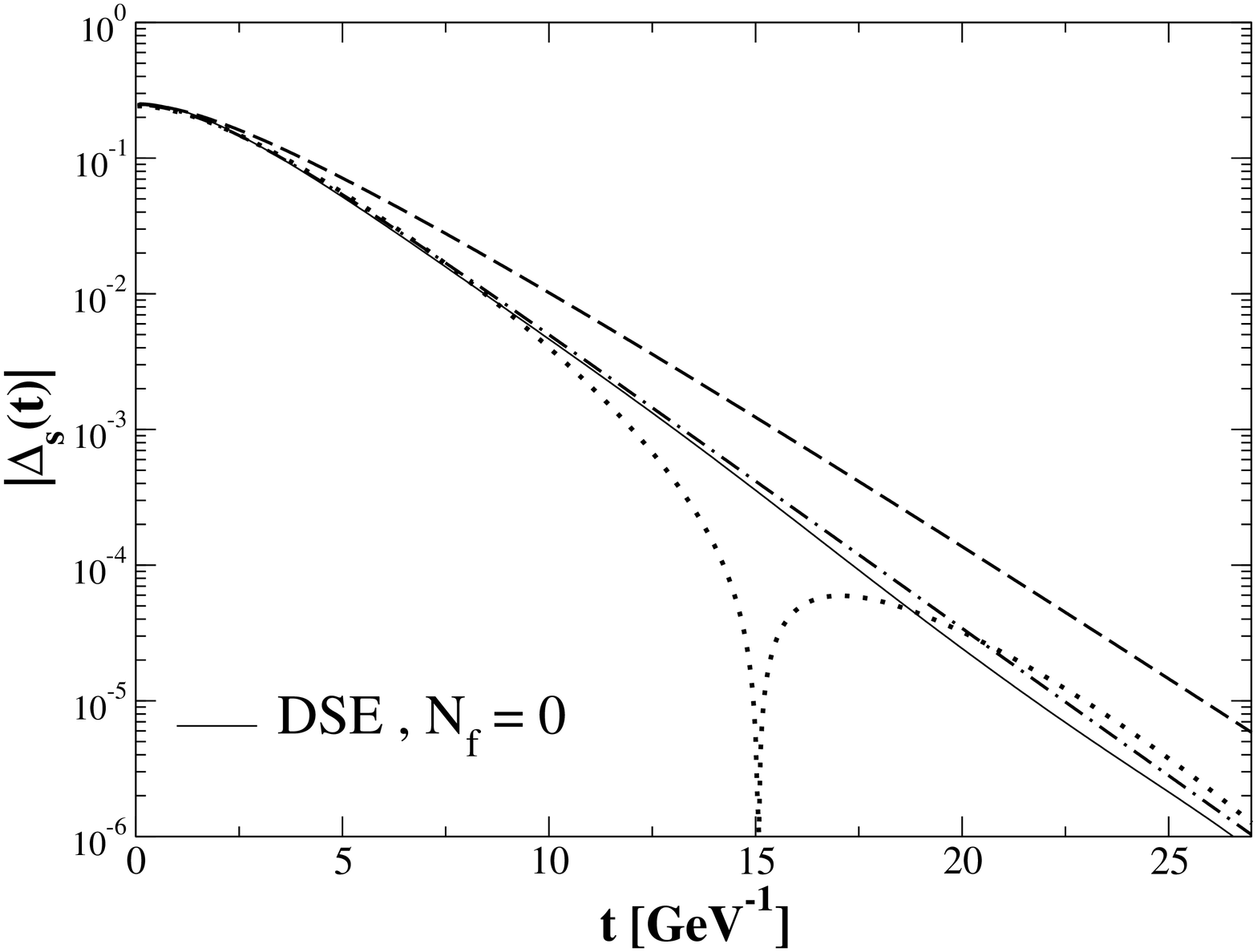,width=80mm,height=65mm}
    \hspace{5mm}
    \epsfig{file=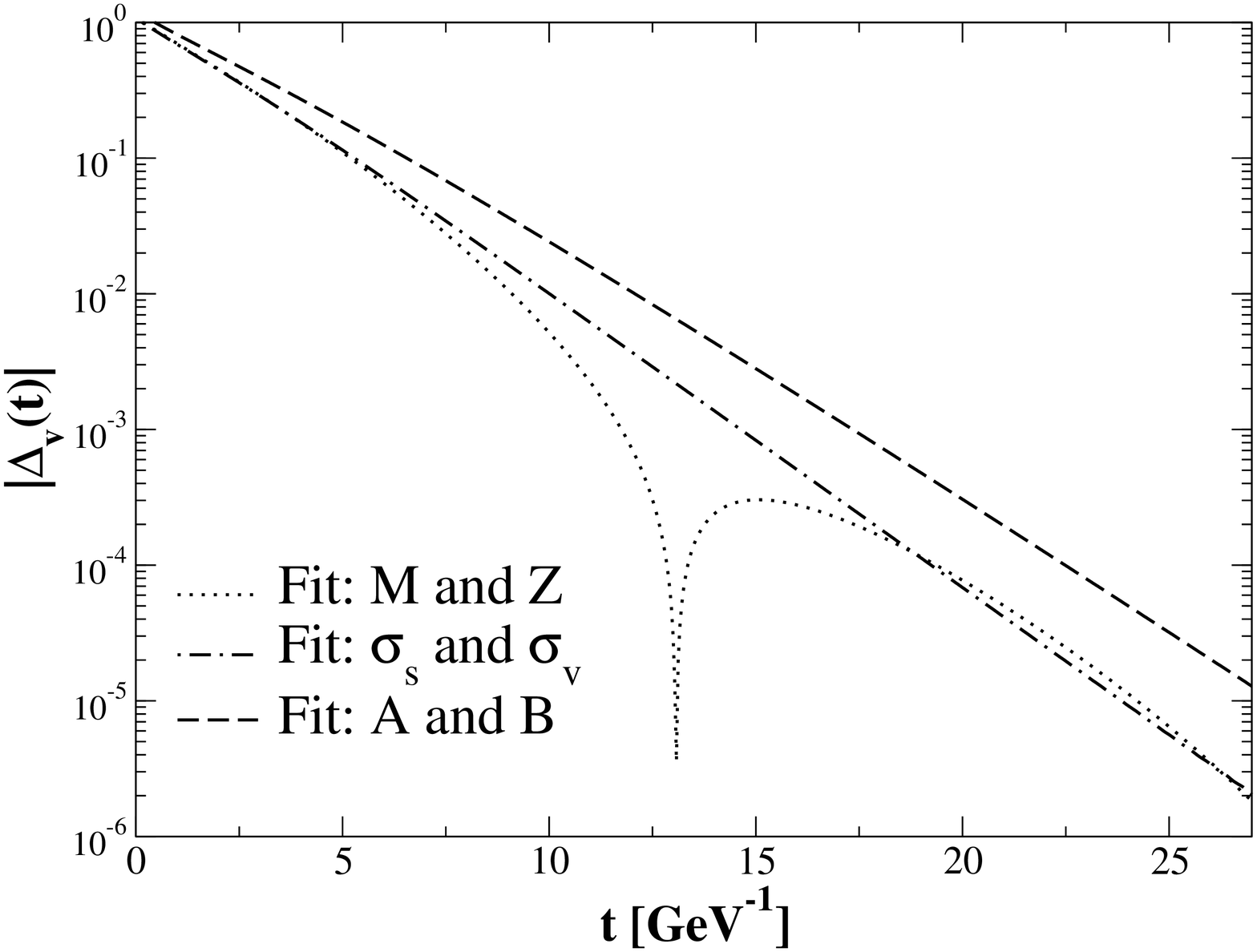,width=80mm,height=65mm}
  }
  \caption{\label{fig:cutfft} The Schwinger functions $\Delta_s(t)$ 
    (left) and $\Delta_v(t)$ (right) for the three parameterizations 
    with branch cuts, in the chiral limit.  For comparison, we have 
    also included our DSE results.}
\end{figure}

Despite the fact that these parameterizations fit $Z^f(p^2)$ and
$M(p^2)$ so well, the corresponding Schwinger functions do not fit the
Schwinger functions of the DSE solutions.  Clearly, the zeros of $p^2
+ M^2(p^2)$ (which determine the poles of $\sigma_{s,v}(p^2)$) will in
general not occur on the negative $p^2$ axis when Eq.~(\ref{eq:fitM})
is used for $M(p^2)$.  Indeed, the dominant singularities of the
propagator functions $\sigma_{s,v}(p^2)$ calculated from the
parameterizations of $Z^f(p^2)$ and $M(p^2)$ are a pair of complex
conjugate singularities, and the corresponding Schwinger functions
clearly show oscillations\footnote{For the heavier quarks, these
oscillations are numerically difficult to detect because the Fourier
transform falls off very rapidly with $t$.} (see
Fig.~\ref{fig:cutfft}).  Extensive ``fine-tuning'' of the fitting form
and/or the parameters is required in order for $p^2 + M^2(p^2)$ to
have its first zero at the pole mass deduced from the Schwinger
function of the DSE solution.

%%%%%%%%%%%%%%%%%%%%%%%%%%%%%%%%%%%%%%%%%%%%%%%%%%%%%%%%%%%%%%%%%%%%%%
%%%%%%%%%%%%%%%%%%%%%%%%%%%%%%%%%%%%%%%%%%%%%%%%%%%%%%%%%%%%%%%%%%%%%%

As a second alternative, we can directly parameterize
$\sigma_{s,v}(p^2)$, and fit these to the numerical DSE solutions.
Again, we want to reproduce the leading logarithmic corrections to
$Z^f(p^2)$ and $M(p^2)$, which can be achieved by using the forms
\begin{eqnarray}
B_{\hbox{\scriptsize chiral}}(p^2) &=& 
        C_{dcsb} \; \frac{\alpha(p^2 + \msing^2)^{1-\gamma_m}}
                         {p^2 + \msing^2 + \Lambda^2}
                  + \frac{C_4}{(p^2 + \msing^2 + \Lambda^2)^2} \,,
\label{eq:fitBchiral}
\\
\sigma_s(p^2) &=& \frac{B_{\hbox{\scriptsize chiral}}(p^2)}{p^2 + \msing^2} +
                  \frac{C_{cqm} \; \alpha(p^2 +
\msing^2)^{\gamma_m}}{p^2 + \msing^2} \,,
\label{eq:fitsigs}
\\
\sigma_v(p^2) &=& \frac{1}{p^2 + \msing^2} \left(1 
                - \frac{\alpha(p^2 + \msing^2)}{2\pi}
                + \tilde{C} \; B_{\hbox{\scriptsize chiral}}(p^2)  
                  \right)  \,.
\label{eq:fitsigv}
\end{eqnarray}
This form has mass-like singularities in $\sigma_{s,v}(p^2)$ at
$p^2=-m_{\rm sing}^2$ from which branch cuts extend to $p^2=-\infty$.
Away from the real axis, $\sigma_{s,v}(p^2)$ have no singularities
(though there is a second singularity at $p^2=-\msing^2-\Lambda^2$).
Furthermore, this parameterization ensures the correct asymptotic
behavior, both for $\sigma_{s,v}(p^2)$ and for the quark functions
$M(p^2)$ and $Z^f(p^2)$.  The main disadvantage of fitting
$\sigma_{s,v}(p^2)$ is that the analytic structure of $Z^f(p^2)$ and
$M(p^2)$, and of $A(p^2)$ and $B(p^2)$ will now become non-trivial.
Again, a delicate fine-tuning is required to obtain a good fit for
both $\sigma_{s,v}(p^2)$ and for $Z^f(p^2)$, $M(p^2)$, $A(p^2)$ and
$B(p^2)$.

\begin{figure}
  \vspace{5mm}\centerline{
    \epsfig{file=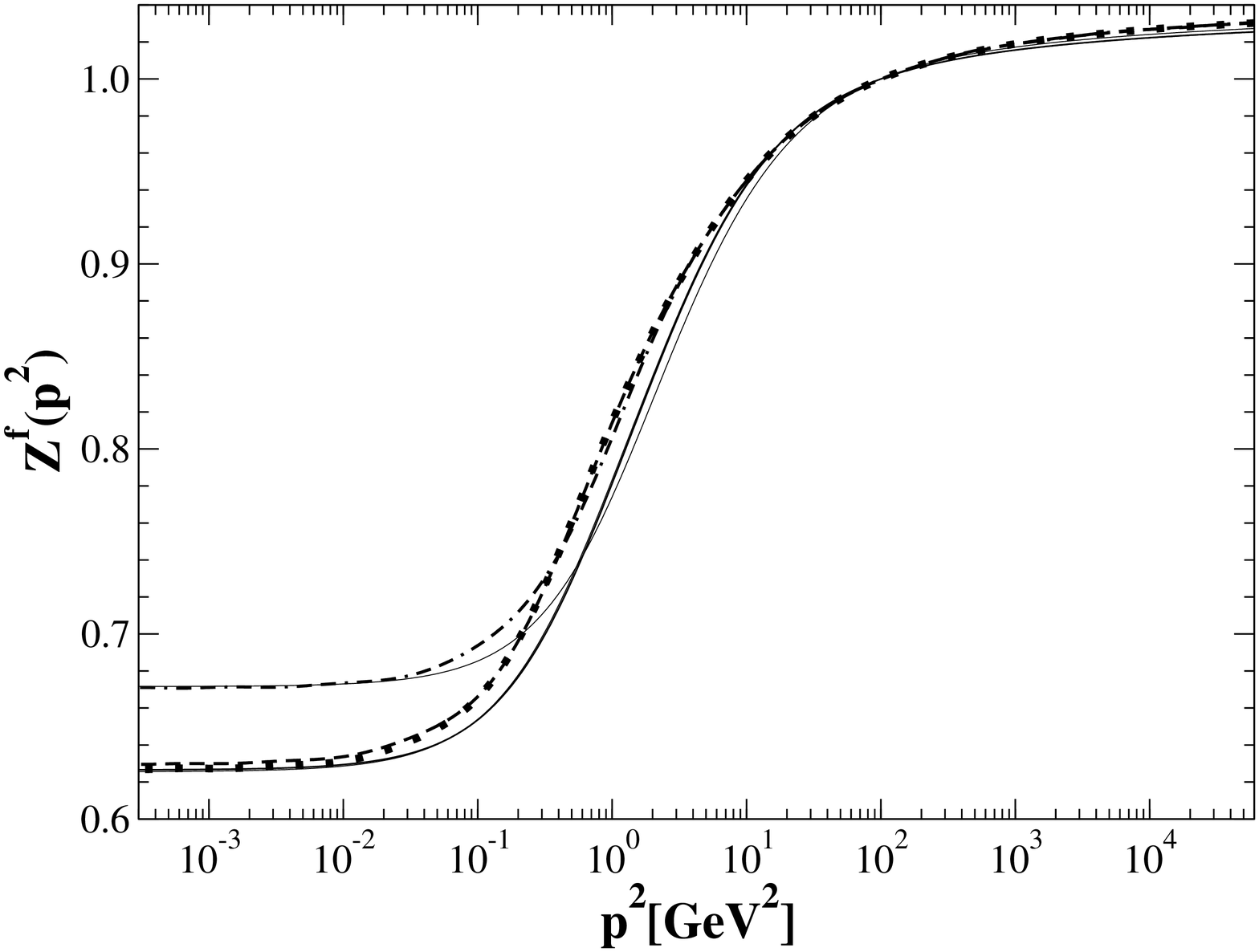,width=80mm,height=65mm}
    \hspace{5mm}
    \epsfig{file=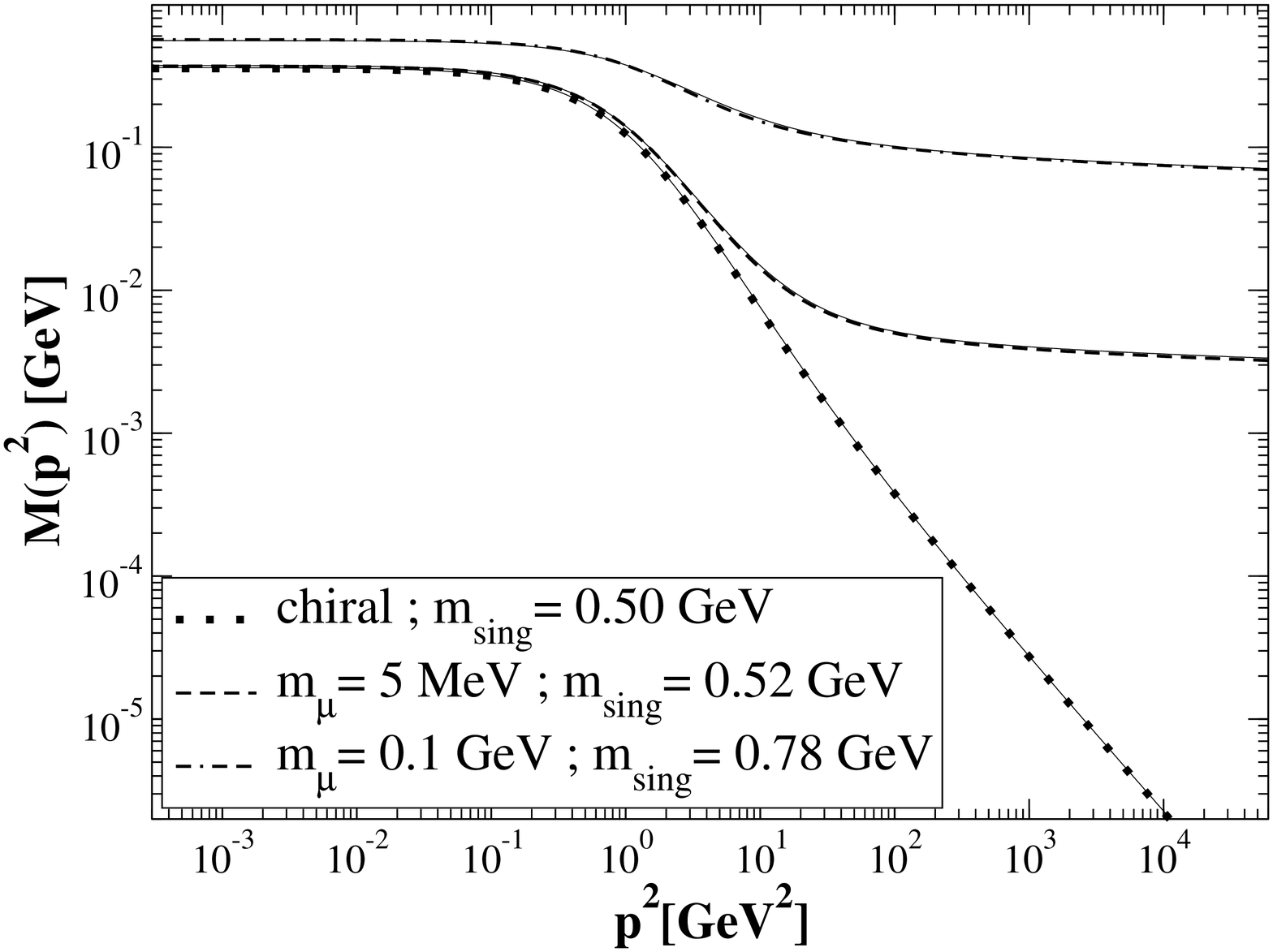,width=80mm,height=65mm}
  }
  \caption{\label{fig:fitDSEsigma}
    The functions $Z^f(p^2)$ (left) and $M(p^2)$ (right) obtained from 
    parameterizations of $\sigma_{s,v}(p^2)$,
    Eqs.~(\ref{eq:fitBchiral})--(\ref{eq:fitsigv}), fitted to the numerical
    DSE solutions for different current quark masses.  The solid
    curves are the fits, with the fit parameters given in 
    Table~\ref{table:parameters}.}
\end{figure}

The parameters play a similar role to those in the previous
parameterization, with the exception of $\tilde{C}$ which is
determined by requiring that $Z^f(p^2)$ is finite at the mass pole.
The other parameters are fixed by fitting $Z^f(p^2)$, $M(p^2)$, and
the Schwinger functions.  For moderately small current quark masses,
we can obtain reasonably good fits, as can be seen in
Fig.~\ref{fig:fitDSEsigma}, with the corresponding parameters listed
in Table~\ref{table:parameters}.  We can also fit the Asqtad lattice
data quite well with this parameterization, as shown in
Fig.~\ref{latcomp}.  For current quark masses larger than a few
hundred MeV, the wave function renormalization can no longer be fitted
with this relatively simple form.  This is most likely related to the
substantial increase in the constant $C_2$ for heavy quarks when
fitting $Z^f(p^2)$ directly (see Table~\ref{table:parameters}).

The functions $A(p^2)$, $B(p^2)$, $M(p^2)$ and $Z^f(p^2)$ have a
singularity at $p^2 = -\msing^2$ where a branch cut along the negative
real axis starts, and another singularity further in the timelike
region at $p^2 = -\msing^2 - \Lambda^2$.  In addition, $M(p^2)$ and
$Z^f(p^2)$ have a pair of complex conjugate poles located at the zeros
of $\sigma_v(p^2)$, and $A(p^2)$ and $B(p^2)$ have two pairs of
complex conjugate poles at zeros of
$p^2\,\sigma_v^2(p^2)+\sigma_s^2(p^2)$.

The Schwinger functions, $\Delta_{s,v}(t)$, are reproduced very well,
see Fig.~\ref{fig:cutfft}.  Notice that the parameterizations of
$Z^f(p^2)$ and $M(p^2)$ fit the DSE solutions for $Z^f(p^2)$ and
$M(p^2)$ better than these parameterizations of $\sigma_{s,v}(p^2)$,
whilst the latter parameterizations are obviously better fits of the
Schwinger functions corresponding to those same DSE solutions.  Thus
we are warned that even an almost perfect fit for Euclidean momenta
does not guarantee a good fit of its Fourier transform, let alone a
good representation of the function in the entire complex plane.

%%%%%%%%%%%%%%%%%%%%%%%%%%%%%%%%%%%%%%%%%%%%%%%%%%%%%%%%%%%%%%%%%%%%%%
%%%%%%%%%%%%%%%%%%%%%%%%%%%%%%%%%%%%%%%%%%%%%%%%%%%%%%%%%%%%%%%%%%%%%%

\begin{figure}
  \vspace{5mm}\centerline{
    \epsfig{file=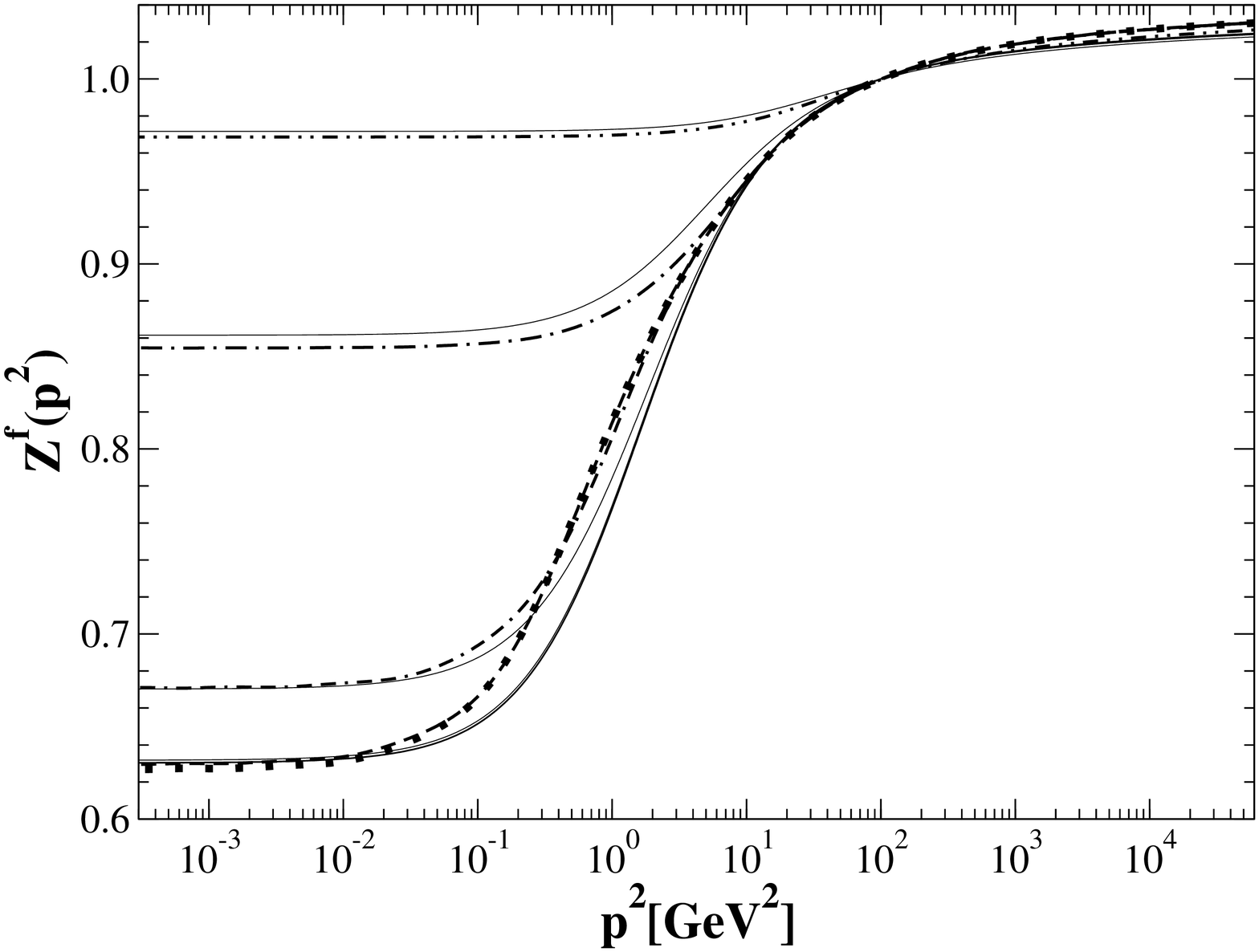,width=80mm,height=65mm}
    \hspace{5mm}
    \epsfig{file=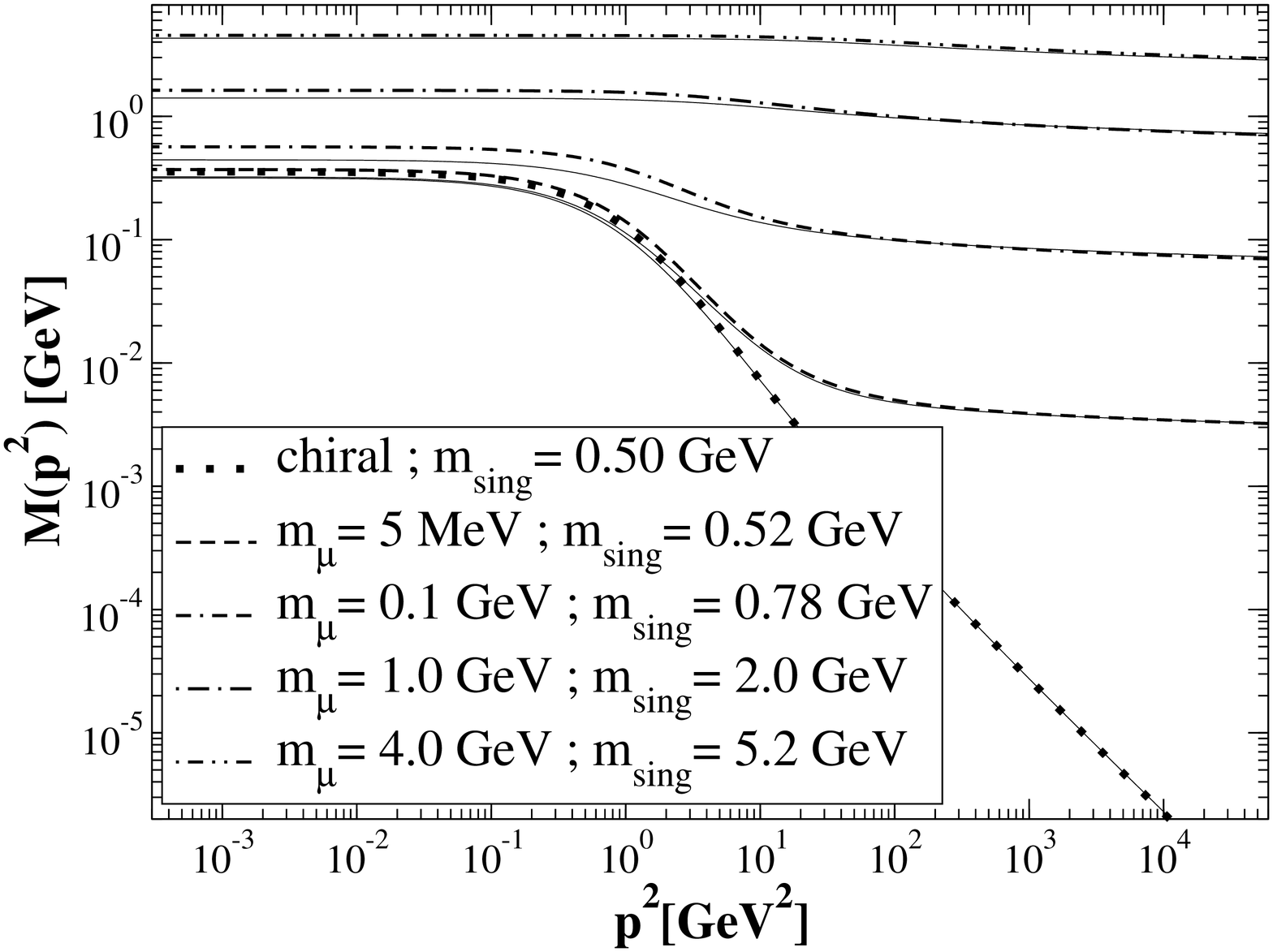,width=80mm,height=65mm}
  }
  \caption{\label{fig:fitDSEAB}
    The functions $Z^f(p^2)$ (left) and $M(p^2)$ (right) obtained from 
    parameterizations of $A(p^2)$ and $B(p^2)$,
    Eqs.~(\ref{eq:fitA})--(\ref{eq:mconstraint}), fitted to the numerical 
    DSE solutions for different current quark masses.  The solid 
    curves are the fits, with the fit 
    parameters given in Table~\ref{table:parameters}.}
\end{figure}

Finally, we construct a parameterization of the inverse quark
propagator functions $A(p^2)$ and $B(p^2)$, such that the propagator
functions $\sigma_{s,v}(p^2)$ have pole-like singularities on the
timelike $p^2$ axis.  For this purpose we use the parameterization
\begin{eqnarray}
Z_2 \; A(p^2) &=& 1 + \frac{\alpha(p^2+\msing^2)}{2\pi} 
            + \frac{C_2}{p^2 + \msing^2 + \Lambda^2}  \;,
\label{eq:fitA}
\\
Z_2 \; B(p^2) &=& C_{dcsb} \; 
    \frac{\alpha(p^2 + \msing^2)^{1-\gamma_m}}{p^2 + \msing^2 + \Lambda^2}
    + \frac{C_4}{(p^2 + \msing^2 + \Lambda^2)^2}  
    + C_{cqm} \; \alpha(p^2 + \msing^2)^{\gamma_m} \,,
\label{eq:fitB} 
\end{eqnarray}
with
\begin{eqnarray}
  \msing &=& \left(C_{dcsb} \; \frac{\alpha(0)^{1-\gamma_m}}{\Lambda^2}
    + \frac{C_4}{\Lambda^4} + C_{cqm} \; \alpha(0)^{\gamma_m}  \right)
   \bigg/\left(1 + \frac{\alpha(0)}{2\pi} + \frac{C_2}{\Lambda^2} \right) \;,
\label{eq:mconstraint}
\end{eqnarray}
and $Z_2$ determined by the renormalization condition $A(\mu^2)=1$.
The results of these fits are shown in Fig.~\ref{fig:fitDSEAB}, with
the corresponding parameters given in Table \ref{table:parameters}.
Though not as good as the direct fits of $Z^f(p^2)$ and $M(p^2)$
(Eqs.~(\ref{eq:fitZ}) and (\ref{eq:fitM})) that were made without
taking into consideration the analytic structure of
$\sigma_{s,v}(p^2)$, these fits reproduce the DSE results within about
10 to 20\% over a wide range of masses.  By construction, the dressing
functions again reduce to the perturbative forms in the ultraviolet
region and the analytic structure is in agreement with the Schwinger
functions corresponding to our DSE solutions.  In the chiral limit,
$\Delta_s(t)$ corresponding to these fits shows significant curvature
at small $t$, as can be seen from Fig.~\ref{fig:cutfft}; for larger
quark masses this curvature decreases.  In contrast, $\Delta_v(t)$ 
does not show this curvature in agreement with our DSE results.
However, the actual analytic structure of $\sigma_{s,v}(p^2)$ is
rather complicated.  In addition to the singularity on the negative
real axis at $p^2 = -\msing^2$ where a branch cut starts, it also has
a pair of complex conjugate poles at zeros of $p^2 \, A^2(p^2) +
B^2(p^2)$.  

%%%%%%%%%%%%%%%%%%%%%%%%%%%%%%%%%%%%%%%%%%%%%%%%%%%%%%%%%%%%%%%%%%%%%%
%%%%%%%%%%%%%%%%%%%%%%%%%%%%%%%%%%%%%%%%%%%%%%%%%%%%%%%%%%%%%%%%%%%%%%

\subsection{Generic features of the quark propagator}

{\em The results of this section point strongly towards an analytical
structure of the quark propagator with a dominant singularity on the
real timelike axis.}  At present, the nature of this singularity
cannot be determined with confidence.  It could be a simple pole, in
which case additional poles or other types of singularities further
away from $p^2=0$ are needed in order to explain the observed
behavior of the Schwinger function.  However, given the structure of
the quark DSE, it is more likely that this singularity is a branch
point, and that there is a branch cut along the negative real axis
starting there.  Having only one branch point singularity on the
negative real axis is (in principle) sufficient to reproduce the
observed Schwinger functions.  We have not yet been able to
distinguish between these alternatives by numerical calculations of
the Schwinger functions of the DSE solutions.

Given the strong sensitivity of the Schwinger functions to the details
of the propagator functions, and the fact that the dominant
singularity is well into the timelike region, it is unlikely that the
sub-dominant analytic structure of the quark propagator will be
determined by Euclidean lattice simulations.  The situation for the
gluon propagator is quite different: there, the analytic structure is
highly constrained by the behavior for $p^2 \to 0^+$.  By approaching
the singularity at $p^2=0$ from the spacelike region, we can gain
information about the nature of this singularity.  In contrast, the
first non-analytic point of the quark propagator is (most likely) at
$p^2 = -\msing^2 < 0$.  Thus, the behavior of $M(p^2)$ and
$Z^f(p^2)$, or $\sigma_{s,v}(p^2)$, for $p^2 \to 0^+$ does not reveal
much about the analytic properties of the propagator.  We cannot
approach the singularity without accessing the timelike region
explicitly.  In addition to this dominant singularity on (or very
close to) the negative $p^2$ axis, the propagator may have other
sub-leading singularities further away from $p^2 = 0$.  Within the DSE
framework one would have to solve the quark DSE over a suitable region
of the complex momentum plane to decide questions about the nature of
the dominant singularity and about the existence of sub-leading
singularities further from $p^2=0$.  However, this is numerically very
demanding and not within the scope of the present investigation.  As
we have seen in the previous section, the results could also be
strongly influenced by the truncation of the DSEs.

In all of the parameterizations of the preceding subsections a robust
feature appears: the leading singularity is on (or {\it very} near to)
the real axis.  The scale at which this singularity occurs is somewhat
dependent on the constraints used in the fits; the lattice data
suggest a scale of 350 to 390~MeV, whilst the DSE solutions prefer a
slightly larger value $\sim 500$~MeV.  Despite this slight variation,
these numerical values are somewhat intriguing and hint at a possible
interpretation in terms of a constituent quark mass.

Fits like the examples presented here might be usefully applied in
hadron phenomenology pending a more conclusive investigation of the
analytic properties of the quark propagators.  However, one should
treat these parameterizations with care and keep in mind that neither
the fitting forms nor the parameters are unique.  The same Euclidean
data, from lattice or DSE calculations, can be fitted quite well with
different parameterizations having distinct analytic properties.  The
only robust feature concerning the analytic structure is that the
dominant singularity, as probed by the Schwinger function, is on (or
very near) the real timelike axis.

\section{Summary and Conclusion} 

In this work we have investigated the analytic structure of the
propagators of Landau gauge QCD in the timelike momentum region using
Schwinger functions, and employing various analytic parameterizations.
We summarize the main results below.

Both lattice simulations and Dyson-Schwinger equation calculations suggest
that the gluon propagator is finite or even vanishes in the infrared.
The latter behavior necessarily leads to violations of reflection
positivity, a sufficient (but not necessary) condition for gluon
confinement.  Our numerical analysis of the Schwinger functions
calculated from the DSE solutions confirms this behavior, finding
clear evidence of such positivity violations in the gluon propagator
in accordance with previous results
\cite{vonSmekal:1997is,Langfeld:2002dd,Fischer:2003rp}.  The gluon
Schwinger function possesses one zero at $t \sim 1\;{\rm fm}$, marking
the length scale above which sizable negative norm contributions
appear.  We explore the detailed analytic structure of the gluon
propagator in the timelike ($p^2<0$) momentum region by constructing
parameterizations that fit both the spacelike momentum behavior of
the lattice calculations and DSE solutions, and the corresponding
Schwinger function.  These parameterizations incorporate the power-law
infrared behavior determined analytically from the coupled ghost and
gluon DSEs, and the perturbatively calculable ultraviolet logarithmic
behavior.  The crucial feature of these parameterizations is the
presence of a branch cut on the timelike momentum half-axis which
produces the observed positivity violations.  These simple
parameterizations depend on (effectively) only one parameter, the
scale $\Lambda_{QCD}$.

In exploring the analytic structure of the quark propagator using the
same Schwinger function methods, we have found an unexpected
sensitivity of this structure to the truncation of the quark DSE.
Gauge symmetry (or, more precisely, the relevant Slavnov--Taylor
identity) requires the presence of a scalar coupling in the
non-perturbative quark-gluon vertex.  This coupling is only present if
chiral symmetry is broken dynamically, in which case it leads to a
self-consistent enhancement of the effective quark-gluon interaction.
This term is important in at least two ways.  First, it is required in
order that the solution of the coupled quark, gluon, and ghost DSEs
generates enough dynamical chiral symmetry breaking in the quark
propagator to give a phenomenologically acceptable quark condensate
\cite{Fischer:2003rp}.  More importantly for our investigations, the
scalar coupling leads to qualitative changes in the analytic structure
of the quark propagator, and hence has significant consequences for
the questions of (non-)positivity and the manifestation of
confinement.  When this term is omitted (as in the commonly used
rainbow truncation), positivity violations consistent with complex
conjugate singularities in the quark propagator (as found in previous
studies
\cite{Bhagwat:2003vw,Stainsby:1990fh,Maris:1991cb,Bender:1996bm}) are
unambiguously observed.  However, when the gauge-mandated scalar
coupling is included, no such evidence of positivity violation is
found and the dominant analytic structure appears to be a singularity
on the real, timelike ($p^2<0$) axis.  Whilst the absence of
positivity violations says nothing about quark confinement (positivity
violation is a sufficient but not necessary condition), it 
does tell us that confinement is probably not manifest at the level
of the propagator.  We also see similar behavior in (quenched)
QED$_4$.  Here, a positive definite propagator is desirable as the
electron is an observable particle.

Finally in Sec.~\ref{sec:quarkparam}, we have attempted to probe
deeper into the analytic structure of the quark propagator.  We have
constructed various parameterizations and used lattice data, DSE
solutions, and other general properties to constrain them.  Whilst an
infinite variety of functional forms (we have investigated only a few
that come easily to mind -- constructed from real or complex conjugate
poles, and branch cuts on the timelike momentum axis) would be capable
of satisfying our constraint criteria, one robust feature emerges from
our analysis: the dominant ({\it i.e.}, closest to $p^2=0$) analytic
structure occurs on (or  very near to) the real, timelike
half-axis.  The scale of this mass-like singularity, as suggested by
meromorphic parameterizations of the lattice data, is 350 to 390 MeV.
The DSE solution indicates a scale of about 500 MeV.  An accurate
determination of the precise nature of this singularity and additional
sub-dominant contributions awaits future improvements.

\goodbreak

\section*{Acknowledgments}
We are grateful to P.~van Baal, P.~Bowman, H.~Gies, C.~Roberts,
D.~Shirkov, I.~Solovtsov, O.~Solovtsova, P.~Tandy, 
A.~Williams, J.~B.~Zhang, and D.~Zwanziger for many helpful hints and
enlightening discussions.
W.D.\ thanks the members of the Institute for Theoretical Physics 
of the University of T\"ubingen for their hospitality during his visits.
We thank the ECT* for the support of the workshop ``Aspects of Confinement and
Non-perturbative QCD'' in March 2003  at Trento where the research presented
here has been outlined.
This work has been supported by the DFG under contracts Al 279/3-3, 
Al 279/3-4, Gi 328/1-2 and GRK683 (European graduate school
Basel--T\"ubingen) and by the US Department of Energy contract
DE-FG03-97ER41014, and benefited from computer resources provided 
by the National Energy Research Scientific Computing Center.

\goodbreak
\appendix

\section{Further details of the DSE truncation}

Since the quark-gluon vertex appears to play a significant role in the
analytic properties of the quark propagator, we have explicitly
explored the effects of this part of our truncation scheme in the main
body of the text. Here we examine the various truncations we use in
the Yang-Mills sector and investigate reasonable modifications to
highlight the extent of both the truncation dependence and the
truncation independence of our results.

In this work, we have used the perturbative ghost-gluon vertex
\begin{equation}
  \Gamma_\mu^{ghost}(q,p)=i q_\mu\,.
\end{equation}
Lerche and von Smekal \cite{Lerche:2002ep} have investigated a large
class of possible structures for this Green's function. In particular
their results show that such variation leads to infrared behaviour of
the gluon and ghost propagators as in Eqs.~(\ref{z-power}) and
(\ref{g-power}) with the exponent $\kappa$ in the range
$0.5<\kappa<0.7$. In this range the essential analytic properties that
we find for the gluon propagator remain unaltered.

For the three-gluon vertex, we use \cite{Fischer:2003zc}
\begin{equation}
\Gamma_{\rho\nu\sigma}(q,p)= \frac{1}{Z_1(\mu,\Lambda)}
\frac{G(q^2)^{(1-a/\delta-2a)}}{Z(q^2)^{(1+a)}}
\frac{G((q-p)^2)^{(1-b/\delta-2b)}}{Z((q-p)^2)^{(1+b)}}
\Gamma^{(0)}_{\rho\nu\sigma}(q,p)
\end{equation}
where $\Gamma^{(0)}_{\rho\nu\sigma}(q,p)$ is the perturbative form,
$\delta=-9N_c/(44N_c-8N_f)$ is the one-loop anomalous dimension of the
ghost propagator, $Z_1$ is the three-gluon vertex renormalisation, and
we fix $a=b=3\delta$.  As discussed in the main text, this form is
chosen to ensure the running coupling, Eq.~(\ref{alpha_def}), is
renormalisation point independent and that the ghost and gluon
propagators have the correct one-loop anomalous dimensions (these
constraints are satisfied for arbitrary values of the parameters $a$
and $b$).

Whilst no systematic study of this vertex truncation exists, in
Ref.~\cite{Fischer:2003zc} one of us has investigated some variation
of the parameters $a$ and $b$. There, the gluon propagator was found
to vanish at $p^2\to0$ independent of the choice of $a$ and $b$, and
our conclusions about the analytic structure of the gluon propagator
remain qualitatively unchanged.  To investigate the dependence of the
gluon fit parameters ($\omega_{I,II}$ and $\Lambda_{QCD}$) on the
three-gluon vertex truncation, we have made fits to the quenched DSE
solutions using Eq.(\ref{fullfit}) for a number of choices of $a$ and
$b$.  The resulting parameter sets are shown in Table
\ref{gluon-param-table} and are seen to vary by 20\%.  With certain
choices of these parameters, one is able to closely match the quenched
lattice data.  However the original ($a=b=3\delta$) truncation gives
results in adequate agreement (see Fig.~\ref{gluon.dat}) with this
data (which itself is not without error bars) and we use it
exclusively in the main text.

\begin{table}
\begin{ruledtabular}
\begin{tabular}{l|cc|cc}
Truncation & \multicolumn{2}{c}{Fit I} & \multicolumn{2}{c}{Fit II} 
\\ \hline
& $\omega_I$ & $\Lambda_{QCD}$ [MeV] & $\omega_{II}$ & $\Lambda_{QCD}$ [MeV]
\\ \hline
Full QCD, $a=3\delta$ & 2.4(2.0) & 500(470) & 2.5 & 510
\\
Quenched QCD, $a=3\delta$& 2.9 & 550 & 3.2 & 550
\\
Quenched QCD, $a=2\delta$ & 2.7 & 410 & 3.0 & 400
\\
Quenched QCD, $a=4\delta$ & 3.2 & 560 & 3.6 & 550
\\
\end{tabular}
\end{ruledtabular}
\caption{\label{gluon-param-table} 
  Gluon fit parameters for various truncations of the system of
  DSEs. The parenthesized results for the full QCD fit I are when the
  fit is optimized to reproduce fit the DSE Schwinger function; all
  other parameter sets are fitted to $Z(p)$.}
\end{table}

Finally, the truncation we employ neglects the effects of the
four-gluon vertex.  These effects are unknown up to now. Since the
ghost loop is dominant in the infrared \cite{Zwanziger:2003cf} and the
one-loop diagrams dominate in the perturbative, ultraviolet region,
such effects are expected to be most important in the intermediate
momentum regime ($p^2\sim 1$GeV$^2$). A two-parameter model for the
corresponding two-loop diagrams in the coupled gluon-ghost DSEs has
been explored in Ref.~\cite{Bloch:2003yu}; under such variation the
$p^2\to0$ behaviour of the gluon and ghost propagator remains
qualitatively the same as in our results.

\section{Pole or branch point in the quark propagator?}

\begin{figure}
  \vspace{5mm}\centerline{
    \epsfig{file=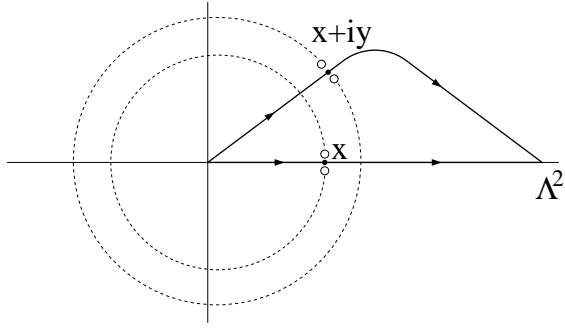,width=80mm}
  }
  \caption{\label{fig:cut}
  Location of the branch cuts (dotted curves) in $K(q^2, p^2)$ in the 
  complex $q^2$-plane, for $p^2=x$ being real and for a complex value  
  of $p^2 = x + i\,y$.  Also shown are possible integration paths 
  (solid curves) from $0$ to $\Lambda^2$ that do not cross the
  corresponding branch cuts.}
\end{figure}

If the quark propagator has a non-analytic point at $p^2 = -\msing^2$
where the propagator diverges, what kind of singularity can we expect?
In order to answer this question, consider the generic integral that
appears in the RHS of the quark DSE
\begin{equation} 
  I(p^2) \; = \; \int^\Lambda d^4q\, \frac{\alpha((q-p)^2)}{(q-p)^2} \;  
    K_\theta(q^2, p^2, q\cdot p) \; \sigma_{s,v}(q^2) \; .
\end{equation} 
Assume that the running coupling, $\alpha(k^2)$, does {\em not} go to
zero like $k^2$ for $k^2 \to 0$.  Furthermore, assume that the kernel
$K_\theta(q^2, p^2, q\cdot p)$ has no singularities.  Thus the only
singularities in the integrand are located at $(q-p)^2=0$ and at $q^2
= -\msing^2$ (coming from the propagator function
$\sigma_{s,v}(q^2)$).  For Euclidean values of $p^2$, we calculate
this integral by performing the angular integral first, followed by
the radial integral
\begin{equation} 
  I(p^2) \; = \; \int_0^{\Lambda^2} q^2 dq^2\, \sigma_{s,v}(q^2) \; 
    K(q^2, p^2) \; ,
\label{eq:radint}
\end{equation} 
with
\begin{equation} 
   K(q^2,p^2) \; = \; \int_0^\pi \sin^2\theta \; d\theta
    \frac{\alpha(q^2-2qp\cos\theta+p^2)}{q^2-2qp\cos\theta+p^2} \;  
    K_\theta(q^2, p^2, qp\cos\theta) \; .
\end{equation} 
This angular integral is well-behaved for any Euclidean value of $q^2$
and $p^2$ as long as the singularity in $\alpha(k^2)/k^2$ is an {\em
integrable} singularity, so let us assume from here on that this is
the case.  If we investigate the analytic properties of $K(q^2, p^2)$
for arbitrary complex values of $q^2$, while keeping $p^2$ real and
positive, we discover that it has a branch cut that can be
characterized by $q^2 = p^2 \exp(i\phi)$, with $0<\phi<2\pi$.  Notice
that if we perform the radial integral along the positive real $q^2$
axis, we do not cross this branch cut: the end-points in $\phi$ are
not included.  This is schematically depicted in Fig.~\ref{fig:cut}.
%%%%%%%%%%%%%%%%%%%%%%%%%%%%%%%%%%%%%%%%%%%%%%%%%%%%%%%%%%%%%%%%%%%%%%
%%%%%%%%%%%%%%%%%%%%%%%%%%%%%%%%%%%%%%%%%%%%%%%%%%%%%%%%%%%%%%%%%%%%%%

Now consider the analytic continuation to complex values of $p^2$.
Clearly, the branch cut in $K(q^2, p^2)$ will move, as indicated in
Fig.~\ref{fig:cut}.  This means that the integration path in
Eq.~(\ref{eq:radint}) has to be deformed so as not to cross the
(shifted) branch cut stemming from the angular integration, while
keeping the end-points fixed.  A possible integration path has been
shown in Fig.~\ref{fig:cut}, though the actual integration path is of
course not unique.  The general rule for this deformed integration
path is that it has to go through the point $q^2 = p^2$, since that is
where there is an opening in the circular branch cut of $K(q^2, p^2)$.
This procedure leads to a well-defined and unambiguous analytic
continuation of the Euclidean DSE, and can be implemented numerically
\cite{Maris:1995ns,pmthesis}.

%%%%%%%%%%%%%%%%%%%%%%%%%%%%%%%%%%%%%%%%%%%%%%%%%%%%%%%%%%%%%%%%%%%%%%
\begin{figure}
  \vspace{5mm}\centerline{
    \epsfig{file=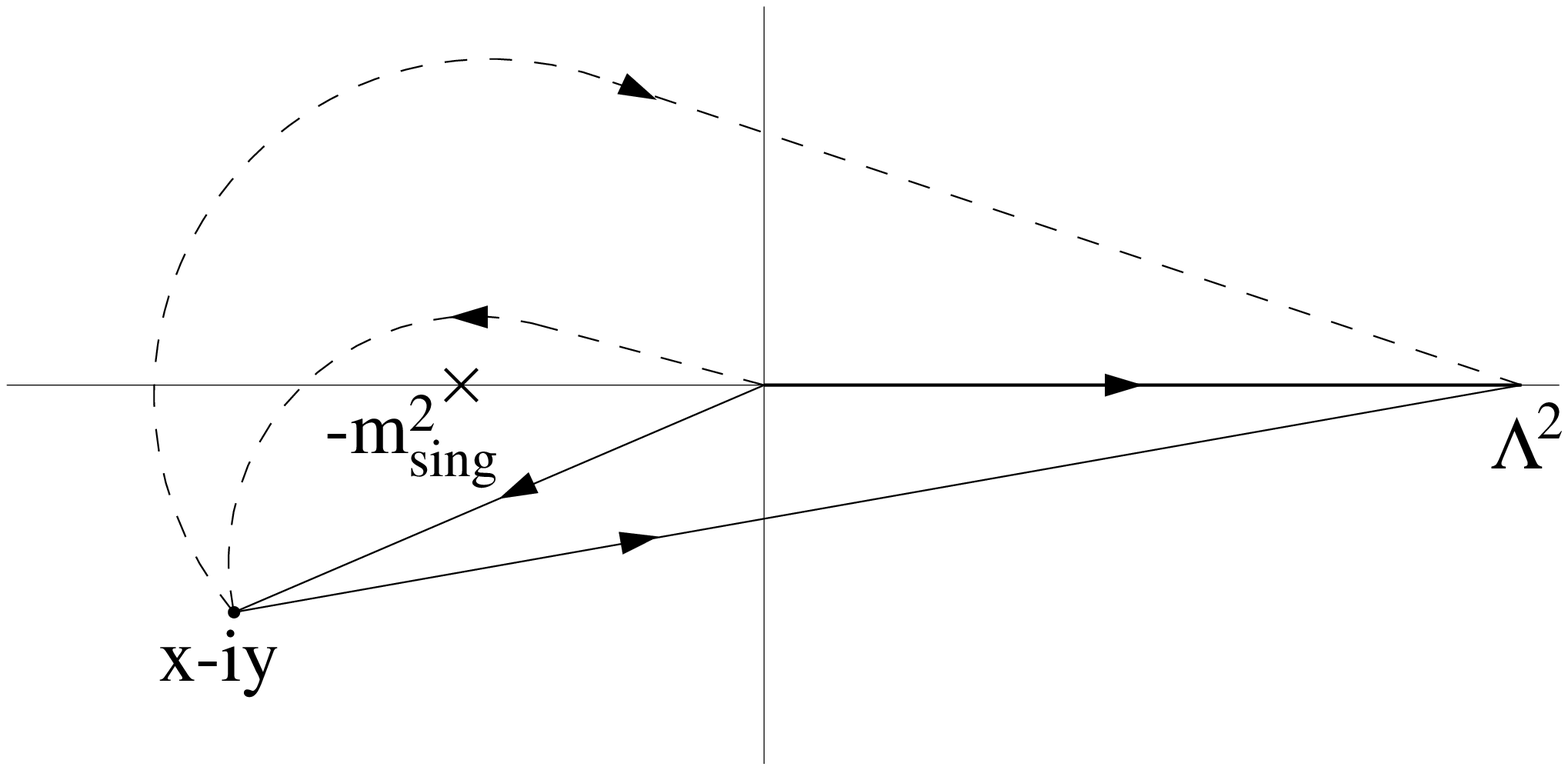,width=80mm}
    \epsfig{file=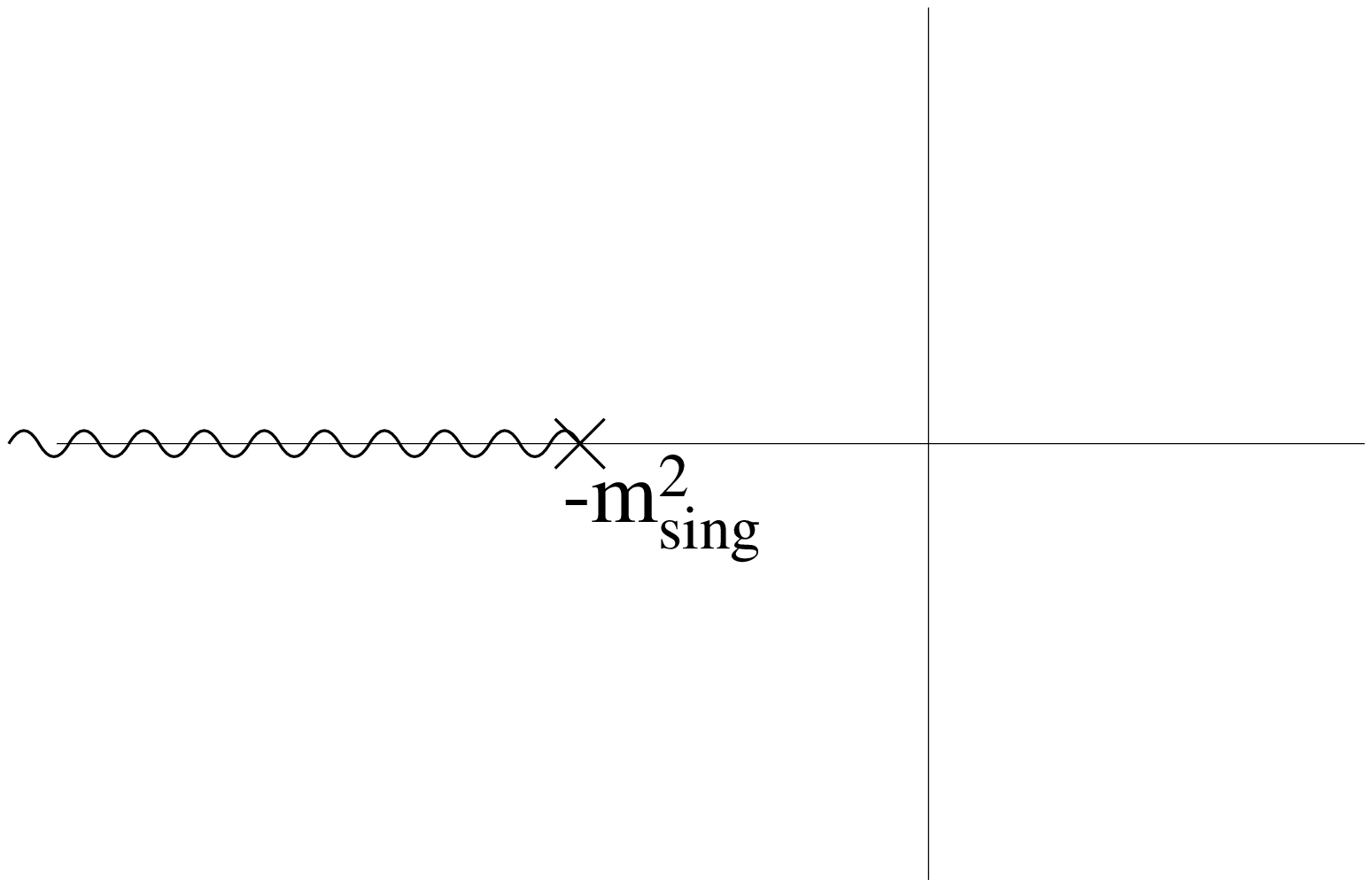,width=80mm}
  }
  \caption{\label{fig:branchpoint}
  Two integration contours in the complex $k^2$-plane for the radial 
  integral in $I(p^2)$ for $p^2=x-iy$ (left) and the resulting 
  analytic structure for $I(p^2)$ in the complex $p^2$-plane (right).}
\end{figure}
%%%%%%%%%%%%%%%%%%%%%%%%%%%%%%%%%%%%%%%%%%%%%%%%%%%%%%%%%%%%%%%%%%%%%%
Following this procedure, one can now show that any singularity in
$\sigma_{s,v}(q^2)$ leads (in general) to a branch point singularity
in $I(p^2)$.  This is shown in detail in Fig.~\ref{fig:branchpoint}
for the case in which $\sigma_{s,v}(q^2)$ have singularities on the
real time-like axis at $q^2=-\msing^2$.  We have drawn two distinct
radial integration paths in order to calculate $I(p^2)$ for $p^2 =
x-iy$. One is obtained by continuously deforming the original
integration path through the upper half of the complex $q^2$-plane,
crossing the negative real axis beyond $q^2=-\msing^2$ (dashed curve),
and the other by deforming the integration path via the lower half of
the complex plane (solid curve).  Because of the combination of (i)
the singularity in $\sigma_{s,v}(q^2)$ at $q^2=-\msing^2$ and (ii) the
circular branch cut in $K(q^2,p^2)$, these two integration paths
cannot be deformed into each other while keeping $p^2$ fixed.
Therefore, the obtained values of $I(p^2)$ will (in general) be
different, and $I(p^2)$ becomes a multi-valued function with a
branch-point singularity at $p^2 = -\msing^2$.  The ``natural'' choice
for the branch cut is along the negative real axis, as indicated by
the wavy line in the right panel of Fig.~\ref{fig:branchpoint}.

Returning to the specific case of the quark propagator, we note that
the RHS of the quark DSE contains an integral like $I(p^2)$, whereas
the LHS is one of the inverse quark propagator functions $A(p^2)$ or
$B(p^2)$.  If $\sigma_{v,s}(q^2)$ has a singular point
$k^2=-\msing^2$, then $I(p^2)$ has a branch-point singularity at
$p^2=-\msing^2$, and therefore $A(p^2)$ and $B(p^2)$ will have a
branch point at $p^2=-\msing^2$.  Thus, unless there are intricate
cancellations, the singularity in $\sigma_{v,s}(p^2)$ of a
self-consistent solution is a branch-point singularity and not a
simple pole.

\goodbreak

\end{document}